\documentclass[twocolumn,aps,pra]{revtex4}

\usepackage{amsmath,amsfonts,amssymb}
\usepackage{wrapfig}
\usepackage{graphicx}
\usepackage{bbm}
\usepackage{color}
\usepackage{float}
\usepackage{subfigure}
\usepackage{textcomp}
\usepackage[english]{babel}

\usepackage{mathrsfs}
\usepackage{wasysym} 
\usepackage{psfrag}
\usepackage[colorlinks=true,linkcolor=blue,citecolor=blue]{hyperref}
\usepackage{enumerate} 

\newcommand{\be}{\begin{equation}}
\newcommand{\ee}{\end{equation}}
\newcommand{\bea}{\begin{eqnarray}}
\newcommand{\eea}{\end{eqnarray}}
\newcommand{\ba}{\begin{array}}
\newcommand{\ea}{\end{array}}
\newcommand{\bc}{\begin{center}}
\newcommand{\ec}{\end{center}}
\newcommand{\ben}{\begin{enumerate}}
\newcommand{\een}{\end{enumerate}}
\newcommand{\bi}{\begin{itemize}}
\newcommand{\ei}{\end{itemize}}
\newcommand{\bt}{\begin{table}}
\newcommand{\et}{\end{table}}
\newcommand{\btab}{\begin{tabular}}
\newcommand{\etab}{\end{tabular}}
\newcommand{\bfi}{\begin{figure}}
\newcommand{\efi}{\end{figure}}

\newcommand{\bd}{\begin{description}}
\newcommand{\ed}{\end{description}}

\newcommand{\lb}{\label}

\newcommand{\tc}{\textcircled}

\makeatletter

\def\compoundrel#1\over#2{\mathpalette\compoundreL{{#1}\over{#2}}}
\def\compoundreL#1#2{\compoundREL#1#2}
\def\compoundREL#1#2\over#3{\mathrel
      {\vcenter{\hbox{$\m@th\buildrel{#1#2}\over{#1#3}$}}}}
\makeatother

\usepackage{babel}
\usepackage{wasysym}
\usepackage{harmony}
\usepackage{semtrans}

\makeatother

\def\be{\begin{eqnarray}}
\def\ee{\end{eqnarray}}
\def\bee{\begin{eqnarray*}}
\def\eee{\end{eqnarray*}}

\begin{document}

\title{Projective simulation for artificial intelligence}

\author{Hans J. Briegel$^{1,2}$ and Gemma De las Cuevas$^{1,2}$}

\affiliation{$^1$Institut f{\"u}r Theoretische Physik,
Universit{\"a}t Innsbruck, Technikerstra{\ss }e 25, A-6020 Innsbruck\\
$^2$Institut f{\"u}r Quantenoptik und Quanteninformation der
\"Osterreichischen Akademie der Wissenschaften, Innsbruck, Austria
}

\date{\today}

\begin{abstract}
We propose a model of a learning agent whose interaction with the environment is governed by a \emph{simulation-based projection}, which allows the agent to project itself into future situations before it takes real action. Projective simulation is based on a random walk through a network of clips, which are elementary patches of episodic memory. The network of clips changes dynamically, both due to new perceptual input and due to certain compositional principles of the simulation process. During simulation, the clips are screened for specific features which trigger factual action of the agent. The scheme is different from other, computational, notions of simulation, and it provides a new element in an embodied cognitive science approach to intelligent action and learning. Our model provides a natural route for generalization to quantum-mechanical operation and connects the fields of reinforcement learning and quantum computation.
\end{abstract}

\maketitle


\section{Introduction}

Computers of various sorts play a role in many processes of modern society. A prominent example is the personal computer which has a specific user interface, waiting for human input and delivering output in a prescribed format. Computers also feature in automated processes, for example in the production lines of a modern factory. Here the input/output interface is usually with other machinery,
such as a robot environment in a car factory.

An increasingly important role is played by so-called \emph{intelligent agents} that operate autonomously in more complex and changing environments. Examples of such environments are traffic, remote space, but also the internet. The design of intelligent agents, specifically for tasks such as learning \cite{SuttonBarto98}, has become a unifying agenda of various branches of artificial intelligence \cite{RusselNorvig03}. Intelligence is hereby defined as the capability of the agent to perceive and act on its environment in a way that maximizes its chances of success. In recent years, the field of embodied cognitive sciences \cite{PfeifferScheier99} has provided a new conceptual and empirical framework for the study of intelligence, both in biological and in artificial entities.

A particular manifestation of intelligence is creativity and it is therefore natural to ask: To what extent can agents or robots show \emph{creative behavior}? Creativity is hereby understood as a distinguished capability of dealing with unprecedented situations and of relating a given situation with other \emph{conceivable} situations. A similar question may arise in behavioral studies with animals, and it is related, on a more fundamental level, to the problem of free will \cite{FreeWill}.

In this paper, we introduce a scheme of information processing for intelligent agents which allows for an element of creative behavior in the above sense.
Its central feature is a \emph{projection simulator (PS)} which allows the agent, based on previous experience --and variations thereof-- to project itself into potential future situations. The PS uses a specific memory system, which we call \emph{episodic \& compositional memory} (ECM) and which provides the platform for simulating future action before real action is taken. The ECM can be described as a stochastic network of so-called \emph{clips}, which constitute the elementary \emph{excitations} of episodic memory \footnote{The notion of episodic memory was introduced by Endel Tulving \cite{Tulving72} in psychology, which was later adopted by cognitive neuroscience. It must be emphasized, however, that our model of ``episodic'' memory is much more primitive and does e.g.\ not assume any encoding of time or the ability of dating experience. The ``clips'' we are introducing could be regarded as primitive forms of episodic memory within a physical toy model.}. Projective simulation consists of a replay of clips representing previous experience, together with the creation of new clips under certain variational and compositional principles. The simulation requires a platform which is detached from direct motor action and on which fictitious action is continuously ``tested''. Learning takes place by a continuous modification of the network of clips, which occurs in three distinct ways: (1) adaptive changes of transition probabilities between existing clips (\emph{bayesian updating}); (2) creation of new clips in the network via new perceptual input (\emph{new clips from new percepts}); (3) creation of new clips from existing ones under certain compositional principles (\emph{new clips through composition}).

In modern physics, the notion of simulation and the ultimate power of physical systems to simulate other systems has become one of the central topics in the field of quantum information and computation \cite{NielsenChuang}. A timely example is the universal quantum simulator, which is capable of mimicking the time evolution of any other quantum system as described by Schr{\"o}dinger's equation of motion; other examples are classical stochastic simulators that mimic the time-evolution of some complex process such as the weather or the climate. These are all examples of \emph{dynamic simulators}, which simulate (that is, compute) the time evolution of a system according to some specified law. It is important to note that these notions of simulators build on \emph{prescribed law}, e.g. certain equations of motion provided by physical, biological, or ecological theory.

The projection simulator that we discuss in this paper -- both its classical and its quantum version -- is entirely different and should be distinguished from these notions of simulators. As in standard theory of reinforcement learning \cite{SuttonBarto98}, our notion of projective simulation builds entirely on \emph{experience} (i.e. previously encountered perceptual input together with the actions of the agent). Projective simulation can be seen, in general terms, as a continuous feedback scheme of a system (agent) endowed with some memory, interacting with its environment. The function of PS is to re-excite fragments of previous experience (clips) to simulate future action, before real action is taken. As part of the simulation process, sequences of \emph{fictitious memory} will be created by a probabilistic excitation process. The contents of these fictitious sequences are evaluated and screened for specific features, leading to specific action. The episodic and compositional memory thereby provides a \emph{reflection and simulation platform} which allows the agent to detach from primary experience and to project itself into \emph{conceivable} situations \footnote{These are situations which are, in memory space, within a certain range of likelihood under the rules of clip variation and composition (see Section \ref{SectionProjectedSimulation}).}.

There is a body of literature in the fields of artificial intelligence and machine learning, where ideas of learning and simulation have been discussed in various contexts (for modern textbook introductions, see e.g. \cite{RusselNorvig03,SuttonBarto98,PfeifferScheier99,FloreanoMattiussi08}). The specific notion of episodic memory and its role for planning and prediction has been discussed in psychology in the 1970s \cite{Tulving72,Ingvar85} and has since been attracting attention in
various fields including cognitive neuroscience and brain research, reinforcement learning, and even robotics \cite{ClarkGrush99,Hesslow02,Schacter08,Hasselmo11,Lin92,Sutton90,SuttonEtAl99,OrmoneitSen02,SuttonEtAl08,Tani96,HoffmannMoeller04,VaughanZuluage06,Toussaint06,ButzEtAl10,Holland75}.
The model which we develop here differs however from previous work in essential respects, as will be elaborated on below.

Our model aims at establishing a general framework that connects the embodied agent research with fundamental notions of physics. This requires a notion of simulation in agents that is both physically grounded and sufficiently general in its constitutive concepts. We claim that the abstract notion of clips and of projective simulation as a random walk through the space of clips provides such a general framework, which allows for different concrete realizations and implementations. This framework also allows us to generalize the model to quantum simulation, thereby connecting the problem of artificial agent design to fundamental concepts in quantum information and computation.

The plan of the article is as follows. In Section~\ref{SectionIntelligentAgents} we briefly review the standard definition of artificial agents. In Section~\ref{SectionProjectedSimulation} we introduce and describe in more detail the projection simulator and our scheme of a learning agent based on episodic \& compositional memory. Section~\ref{SectionFormalDefinitions} introduces some formal notation. Section~\ref{SectionSimpleExamples} provides illustrations of the main concepts using examples of a learning agent in a simple computer game. In Section~\ref{SectionLiterature} we compare our model of projective simulation with some related work in the fields of artificial intelligence, reinforcement learning, and the cognitive sciences.
In Section~\ref{SectionQuantumAgents} we generalize the notion of the projection simulator to a quantum mechanical scheme and discuss the potential role of quantum information processing for artificial agent design. Section~\ref{SectionConclusion} concludes the paper.

\section{Intelligent agents} \label{SectionIntelligentAgents}

\begin{figure}[htb]
\begin{center}
\begin{minipage}{9cm}
\includegraphics[width=8cm]{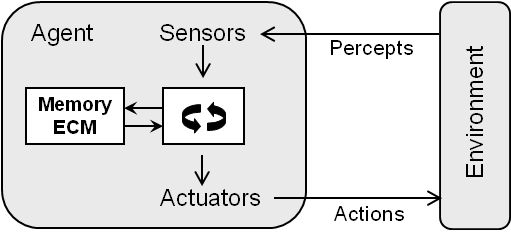}
\end{minipage}
\end{center}
\caption{Model of an agent. Adapted and modified from \cite{RusselNorvig03} (see text).}
\label{AgentModel}
\end{figure}

In the following, we shall discuss the concept of projective simulation in the framework of intelligent agents \cite{RusselNorvig03}. Realizations of intelligent agents could be robots, biological systems, or software packages (internet robots). An agent (see Figure~\ref{AgentModel}) has sensors, through which it perceives its environment, and actuators, through which it acts upon the environment. Internally, one may imagine that it has access to some kind of computing device, on which the agent program is implemented. The function of the agent program is to process the perceptual input and output the result to the actuators.

\noindent For a deterministic agent, a given percept history completely determines the next step (actuator motion) of the agent. For a stochastic agent, it only determines the probabilities with which the agent will perform the possible next actuator moves. In the present paper, we shall deal with the latter situation.

The heart of the agent is usually considered to be its program. The program will depend on the nature of the agent and its environment. It will be different for robots that operate in city traffic, on the surface of a planet, or inside a human body. The environment usually has its own rules that need to be taken into account when designing the program: it is governed by the laws of physics or biology, and it may have limited accessibility, observability, and predictability. The role of the program is to deal with environmental data (through its sensors) and let the agent respond to them in a rational way \cite{RusselNorvig03}.

From a computer-science oriented perspective, it might seem as if the problem of intelligent agents were a mere software problem, i.e.\ reducible to algorithmic design. From such point of view, the ``intelligence'' of the agent is imported and its capability to react rationally within its environment depends entirely on the designer's ingenuity to anticipate all potential situations that the agent may encounter, and thus to build corresponding rules into the \emph{program}.
However, more recent developments in the area of \emph{embodied cognitive science} \cite{PfeifferScheier99} have emphasized physical aspects of the emergence of intelligence, among them the fact that most biological or robotic agents are ``embodied'' and ``situated'', meaning that they acquire information about their environment -- and thereby develop intelligent behavior -- exclusively through physical interactions (via sensors) with the environment.

In this paper, we will adopt such an embodied approach to understanding intelligence \cite{PfeifferScheier99}. We shall concentrate on a specific aspect of intelligence and investigate the possibility of creative behavior in robots or agents. In the spirit of the celebrated work of Braitenberg and his \emph{vehicles} \cite{Braitenberg86}, we will propose an explicit model of memory, which, together with the idea of projective simulation, can give rise to a well-defined notion of creative behavior. The description of episodic memory, as a dynamic network of clips which \emph{grows as the agent interacts} with the world, is thereby fully embedded in the agent architecture.

\section{Learning based on projective simulation} \label{SectionProjectedSimulation}

In this section, we shall focus on one crucial element of the agent architecture, which is its \emph{memory}, indicated by the two connected white boxes in Figure \ref{AgentModel}. There are various and different aspects of memory, which enter in the discussion and which should be kept apart. Research in behavioral neuroscience \cite{KandelNobelPrize} has shown that learning can be related to structural changes on the molecular level of a neural network, providing examples of Hebbian learning \cite{KandelAntonov03}. The behavior of simple animals (such as the sea slug Aplysia \cite{KandelAntonov03}) can largely be described by a stimulus-reflex circuit, where the structure of this circuit changes over time. In the language of artificial agent research, this could be modeled as a reflex agent, whose program is modified over time (which represents the learning of the animal). In such type of learning,  we have a separation of time scales into ``learning'' (shaping of circuit) versus ``reflex'' (execution of circuit) which is possible only for simple agents, but it cannot explain more complex patterns of behavior.

Phenomenologically speaking, more complex behavior seems to arise when an agent is able to ``think for a while'' before it ``decides what to do next.'' This means the agent somehow evaluates a given situation in the light of previous experience, whereby the type of evaluation is different from the execution of a simple reflex circuit. An essential step towards such more complex behavior seems to be the capability of reinvoking memory without inducing immediate motor action, which requires a separate level of representation and storage of previous experience. Such type of memory must thus be decoupled from immediate motor action and cannot, per definition, be part of a reflex circuit.

To model intelligent behavior, people have studied artificial agents of various sorts (utility-based, goal-oriented, logic-based, planning,...) \cite{RusselNorvig03} whose actions are the result of some program or set of rules. In so-called learning agents, the emphasis lies on modeling the \emph{emergence} of behavior patterns when there are no specific rules \emph{a priori} specified, except that the agent remembers in one way or the other that certain percept-action pairs were rewarded or punished (reinforcement learning).

Here we introduce a learning-type agent, whose decisions -- i.e. ``what to do next'' in a given situation -- depend not only on its previous experience with similar situations, but also on \emph{fictitious experience} which it is able to generate on its own. The central element is a projection simulator (PS), together with a type of episodic memory system (ECM), which helps the agent to project itself into ``conceivable'' situations. Triggered by perceptual input, the PS calls memory and induces a random walk through episodic memory space. This random walk is primarily a replay of past experience associated with the perceptual input, which is evaluated before it leads to concrete action. However, memory itself is changed dynamically, both due to actual experience and due to certain compositional principles of memory recall, which may create new content corresponding to fictitious experience that never really happened. In this model, it is essential to have a representation of the environment in terms of the episodic memory, which enables the agent to decouple from immediate connection with the environment and reflect upon its future actions. Importantly, this reflection is not realized as a sophisticated computational process, but it can be seen as a structural-dynamical feature of memory itself.

As a physical basis of the PS, one can imagine a neural-network-type structure, where any primary experience is accompanied by a certain spatiotemporal excitation pattern of the network. The details of this architecture, including the way of encoding information, the concise learning rules, etc., are not important. The only relevant feature is that a later re-excitation with a similar pattern, due to whatever cause, will invoke similar experience.  As the agent learns, it will relate new input with existing memory and thereby change the structure of the network. The only relevant aspect of the neural-network idea is, for our purposes, that any recall of memory is understood as a dynamic re-play of an excitation pattern, which gives rise to episodic sequences of memory.

By episodes we mean patches of stored previous experience. In the specific context of vision, one could also call it a ``movie fragment'' or ``clip''. In the following, we will use the terms \emph{episode} and \emph{clip} interchangeably. Clips represent basic (but variable) units of memory which will be accessed, manipulated, and created by the agent. Clips themselves may be composed of more basic elements of cognition such as color, shape, or motion, but they represent the functional units in our theory of memory-driven behavior.

\begin{figure}[htb]
\begin{center}
\begin{minipage}{9cm}
\includegraphics[width=9cm]{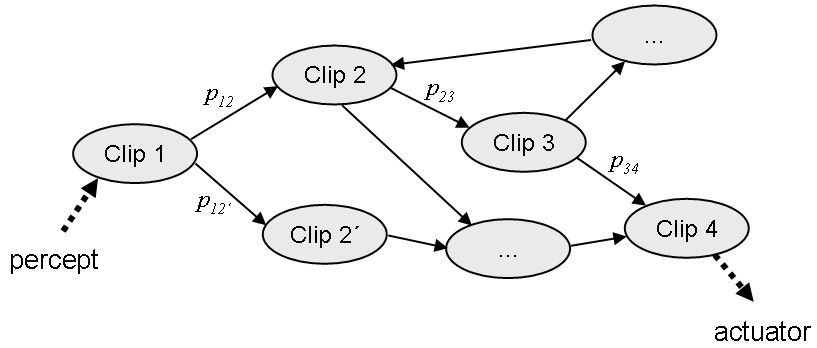}
\end{minipage}
\end{center}
\caption{Model of episodic memory as a network of clips.}
\label{ClipNetwork}
\end{figure}

Formally, episodic memory will be described as a probabilistic network of clips as illustrated in Figure \ref{ClipNetwork}. An excited clip calls, with certain probabilities, another, neighboring clip. The neighborhood of clips is defined by the network structure, and the jump probabilities will be functions of the percept history. In the simplest version, only the jump probabilities (weights) change with time, while the network structure (graph topology) and the clip content is static. In a refined model, new clips (nodes in the graph) may be added, and the content of the clip may be modified (internal dimension of the nodes). \emph{A call of the episodic memory triggers a random walk through this memory space (network). In this sense, the agent jumps through the space of clips, invoking patchwork-like sequences of virtual experience} \footnote{As was mentioned, a clip itself is a composite structure, composed of more basic elements that are drawn from different categories. In case of vision, these categories could be shape, size, motion, color, etc.
Transition probabilities are, to first approximation, defined only between clips. In a refined version, there are also transitions between basic elements, leading to a dynamic modification of clips, as part of the compositional rules of projective simulation.}. Action is induced by screening the clips for specific features. When a certain feature (or combination of features) is present and above a certain intensity level, it will trigger motor action.

In the following sections, we shall put some of these notions in a more formal framework, and illustrate the idea of projective simulation with concrete examples. These examples should be understood as illustrations of the underlying notions and principles. We discuss them in the context of
simple problems of reinforcement learning, but the notion of projective simulation is more general and can be seen as a principle and building block for complete agent architectures.

\section{Mathematical modeling and notation}
\label{SectionFormalDefinitions}

In physical terms, the behavior of an agent (see Figure \ref{AgentModel}) can be described as a stochastic process that maps input variables (percepts) to output variables (actions). An \emph{external view} of the agent consists in specifying, at each time $t$, the conditional probability $P^{(t)}(a|s)$ for action $a\in A$, given that percept $s\in S$ was encountered. This is also called the \emph{agent's policy} in the theory of reinforcement learning \cite{SuttonBarto98}.
Here, $S$ and $A$ denote the set of possible percepts and actuator moves, respectively, which we are going to describe in more detail shortly.

The dependence of this probability distribution on time $t$ indicates, for any non-trivial agent, the existence of \emph{memory} \footnote{Usually, one assumes that the agent operates in cycles, in which case $t$ is an integer variable. When writing $P^{(t)}(a|s)$, one then refers to the conditional probability for choosing action $a=a^{(t)}$ at the end of cycle $t$, if it was presented with $s=s^{(t)}$ at the beginning of the same cycle. In general, the probability with which the agent chooses action $a^{(t)}$ may depend on its entire previous \emph{history}, i.e. the percepts and actions $s^{(t-1)},a^{(t-1)},\dots s^{(1)},a^{(1)}$ in all earlier cycles of the agent's life. However, the interesting part of the agent is how it \emph{learns}, i.e. how its history changes its \emph{internal state}, which in turn determines its future policy. It is the change of its internal state that is indicated by the time-dependence (superscript) of $P^{(t)}(a|s)$, which summarizes the agent's history up to the $t\,$th cycle.}. A corresponding \emph{internal} description connects $P^{(t)}(a|s)$ with the memory of the agent and explains how memory is built up under a given history of percepts and actions.

In our model of the agent, memory consists of a network of episodes (or clips), which are sequences of `remembered' percepts and actions. The operation cycle of an agent can be described as follows: (i) Encounter of percept $s\in S$ which happens with a certain probability $P^{(t)}(s)$ \footnote{The probability $P^{(t)}(s)$ can either be completely independent of the agent's actions, corresponding to an ``open-loop scenario'' or it may depend itself on previous actions of the agent, corresponding to a ``closed loop scenario''.}. The encounter of percept $s\in S$ triggers the excitation of memory clip $c\in C$ according to a fixed ``input-coupler'' probability function ${\cal I}(c|s)$. (ii) Random walk through memory/clip space $C$, which is described by conditional probabilities $p^{(t)}(c'|c)$ of calling/exciting clip $c'$ given that $c$ was excited. (iii) Exit of memory through activation of action $a$, described by a fixed ``output-coupler'' function ${\cal O}(a|c)$.

In the following, we shall only consider finite agents, acting in a finite world. Percepts, actions, and clips are then elements of finite-sized sets, according to the following definitions:

\smallskip\noindent
$\bullet$\emph{Percept space:} \newline
$s \equiv \left(s_1,s_2,\dots ,s_N \right)\in S_1 \times \cdots \times  S_N \equiv S$, $s_i=1,\dots ,|S_i|$.
The structure of the percept space $S$, a cartesian product of sets, reflects the compositional (categorical) structure of percepts (objects). For example, $s_1$ could label the category of shape, $s_2$ category of color, $s_3$ category of size, etc. The maximum number of distinguishable input states is given by the product  $|S|=|S_1|\cdots |S_N|$.

\smallskip\noindent
$\bullet$\emph{Actuator space:} \newline
$a \equiv \left(a_1, a_2,\dots, a_M \right)\in A_1 \times \cdots \times  A_M \equiv A$, $a_j=1,\dots, |A_j|$. The structure of the actuator space $A$ reflects the categories (or, in physics terminology, the degrees of freedom) of the agent's actions. For example $a_1$ could label the state of motion, $a_2$ the state of a shutter, $a_3$ the state of a warning signal, etc. All of this depends on the specification of the agent and the environment. The maximum number of different possible actions is given by the product  $|A|=|A_1|\cdots |A_M|$.

\medskip
Clips or episodes are elementary, short-time, dynamic processes in the agent's memory that relate to past experience and that can be triggered by similar experience. A clip can be seen as a sequence of remembered (real or fictitious) percepts and actions. We distinguish percept $s\in S$ that is \emph{directly} caused by the environment at a given time $t$, from a \emph{remembered }(or a fictitious) percept $\mu(s)\in \mu(S)$ that has a certain representation in the agent's memory system. Similarly, we distinguish \emph{real} actions $a\in A$ executed by the agents from \emph{remembered} (or fictitious) actions $\mu(a)\in \mu(A)$, which can be (re-)called by the agent without necessarily leading to real action. Instead of the symbol $\mu(a)$ we will also use \tc{$a$}$\equiv$ $\mu(a)$ for a remembered action. The formal definition of a clip reads then as follows:

\smallskip\noindent
$\bullet$\emph{Clip space:} \newline
$c\equiv\left(c^{(1)}, c^{(2)}, \dots, c^{(L)}\right)\in C; c^{(l)}\in \mu(S)\bigcup \mu(A)$.
The index $L$ specifies the length of the clip. A simple example for $L=2$
is the clip $c=\left(\mu(s),\mu(a)\right)\equiv $(\tc{$s$},\tc{$a$}), which corresponds to a simple percept-action pair. Clips of length $L=1$ 
consist of a single remembered percept or action, respectively. In the subsequent examples, we will mainly consider probabilistic networks of such simple clips.

Projective simulation is realized as a random walk in episodic memory, which serves the agent to reinvoke past experience and to compose fictitious experience before real action is taken. Learning is achieved by \emph{evaluating} past experience, for example by simple reinforcement learning. In memory, this will lead to a modification of the transition probabilities between different clips, e.g. via Bayesian updating.
We emphasize, again, that such kind of the evaluation happens entirely within memory space. If a certain percept-action sequence $s \to a$ was rewarded at time step $t$, it will typically mean that, in the subsequent time step $t+1$, the transition probability $p^{(t+1)}(a|s)$ between \emph{clips} \tc{$s$} and \tc{$a$} will be enhanced. This is only indirectly related to the conditional probability $P^{(t+1)}(a|s)$ for \emph{real} action $a$ given percept $s$.

For convenience, and to emphasize the role of fictitious experience in episodic memory, we shall also introduce a third space which we call \newline
$\bullet$\emph{Emotion space:} \newline
$e \equiv\left(e_1,e_2,\dots, e_K \right)\in E_1 \times \cdots \times  E_K \equiv E$, $e_k=1,\dots, |E_k|$. In the simplest case $K=1$ and $|E_1|=2$, with a two-valued emotion state  $e_1\equiv e \in \{\smiley, \frownie\}$. Emotional states are
\emph{tags}, attached to transitions between different clips in the episodic memory. The state of these tags can be changed through feedback (e.g. reward) from the environment. They are internal parameters and should be distinguished from the reward function itself, which is defined externally. Informally speaking, emotional states are \emph{remembered rewards} for previous actions, they have thus a similar status as the clips.

The reward function $\Lambda$ is a mapping from $S\times A$ to $I\subset \mathbb{R}$ (real numbers), where in most subsequent examples we consider the case $I={0,1,...,\lambda}$. In the simplest case, $\lambda =1$: If $\Lambda(s,a)=1$ then the transition $s \to a$ is rewarded; if $\Lambda(s,a)=0$, it is not rewarded. A rewarded (unrewarded) transition will set certain emotion tags in the episodic memory to $\smiley$ ($\frownie$), as discussed previously. We shall also consider situations where the externally defined reward function changes in time, which leads to an adaptation of the flags in the agent's memory.

\section{Simple example: Invasion game}
\label{SectionSimpleExamples}

\begin{figure}[htb]
\begin{center}
\begin{minipage}{9cm}
\includegraphics[width=8cm]{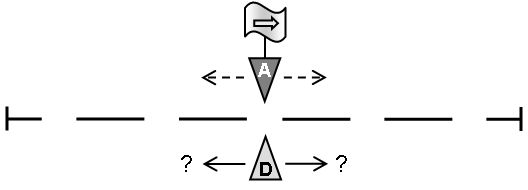}
\end{minipage}
\end{center}
\caption{Game {\em invasion}. Defender agent D, whose task is to block the passage against invasion by the attacker A, tries to guess A's next move from a symbol shown.}
\label{Invasion}
\end{figure}

To illustrate some of these concepts, let us consider the following simple game, which we call {\em invasion} (see Figure \ref{Invasion}). It has two parties, an attacker (A) and a defender (D) (the robot/agent). The task of D is to defend a certain region against invasion by A. The attacker A can enter the region through doors in a wall, which are placed at equal distances. The defender D can block a door and thereby prevent A from invasion.

Initially, defender D and attacker A stand face-to-face at some door $k$, see Figure~\ref{Invasion}. Next, the attacker will move either to the left or to the right, with the intention to pass through one of the adjacent doors. For simplicity, we may imagine that A disappears at door $k$ and re-appears some time $\tau$ later in front of one of the doors $k-1$ or $k+1$. The defender D needs to guess -- based on some information which we will specify shortly -- where A will reappear and move to that door. (We may assume that D moves much faster than A so that, if its guess is correct, it will arrive at the next door before A). If A arrives at an unblocked door, it counts as a successful passage/invasion. The task of D is to hold off the attacker for as long (i.e. for as many moves) as possible. We can define an appropriate blocking efficiency. If A has successfully invaded, this particular duel is over, and the robot D will be faced with a new attacker appearing in front of the door presently occupied by the robot.

Suppose that the attacker A follows a certain strategy, which is unknown to the robot D, but, before each move, A shows some symbol that indicates its next move.
In the simplest case, as illustrated in Figure~\ref{Invasion}, this could be a simple arrow pointing right, $\Rightarrow $, or left, $\Leftarrow$, indicating the direction of the subsequent move. It could also be a whole number, $\pm m$, indicating how far A will move and in which direction \footnote{ We may assume periodic boundary conditions, i.e.\ identify door $N+1$ with door $1$, in which case we need not worry about A hitting a boundary and we need no signs.}. The meaning of the symbols is {\em a priori} completely unknown to the robot, but the symbols can be perceived and distinguished by the robot.  The only requirement we impose at the moment is that the meaning of the symbol stays the same over a sufficiently long period of time (longer than the learning time of the robot). Translated into real life, the ``symbol'' could be as mundane as the ``direction into which the attacker turns it body''  before disappearing (a robot does not know what this means {\em a priori}), it could be an expression on its face, or some abstract symbol that A uses to communicate with subsequent invaders.
The described setup is reminiscent of certain behavior experiments with \emph{drosophila}, using a torsion-based flight simulator system and a reinforcement mechanism to train drosophila to avoid objects in its visual field \cite{MartinHeisenberg,HeisenbergPNAS}. In this sense, the presented analysis may also be interesting for the interpretation of behavior experiments with drosophila or similar species.

Using this simple game, we want to illustrate in the following how the robot can learn, i.e.\ increase its blocking efficiency by projective simulation. We will consider different levels of sophistication of the simulation process (recovering simple reinforcement learning and associative learning as special cases).

Put into the language introduced in the previous section, we consider a percept space that comprises two categories
\hspace*{12pt}\newline$\bullet$
Symbol shown by attacker:  $\{\Leftarrow, \Rightarrow \} = S_1$,
\hspace*{12pt}\newline$\bullet$
Color of symbol: $\{\rm{red}, \rm{blue} \}$ $=S_2$,
\newline while the actuator space comprises a single category
\hspace*{12pt}\newline $\bullet$
Movement of defender: $\{-, +\}=A$,
\newline as does the emotion space
\hspace*{12pt}\newline$\bullet$
Emoticons: $\{\smiley,\frownie\}=E$.

In memory space, \tc{$\Leftarrow $}, \tc{$-$}, etc. correspond to memorized percepts/actions that have been perceived/executed by the agent. In the following, we regard \tc{$\Leftarrow$} and \tc{$-$} as separate clips of length $L=1$.
The role of the emotional tags is to indicate, at a given time, which of the transitions in clip space have recently led to a rewarded action.

For the reward function $\Lambda: S\times A \longrightarrow {0,1,...,\lambda}$, we often consider the simplest case $\lambda =1$ (except where explicitly indicated). For $\Lambda(s,a)=1$ ($0$) the transition $s \to a$ is rewarded (not rewarded). A rewarded transition, $\Lambda(s,a)=1$, will set certain emotion tags in the episodic memory to $\smiley$, which will influence the simulation dynamics. We shall also consider situations where the attacker changes its strategy in time, which leads to a time-dependent reward function and a corresponding adaptation of the flags in the agent's memory.

The conditional probability that a running (or active) clip \tc{$\Leftarrow$} calls clip \tc{$-$} will be denoted by $p^{(n)}(-|\Leftarrow)$, where the upper index $n$ indicates the time step (``experience of the agent''), i.e. how many encounters with an attacker have occured.

\begin{figure}[htb]
\includegraphics[width=5cm]{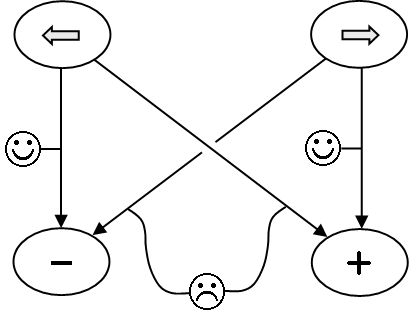}
\caption{Episodic memory that is build up by the defender-agent in Figure \ref{Invasion}, if the attacker follows the static strategy to move one door to the left (right) after showing the symbol $\Leftarrow$ ($\Rightarrow$). The ``emotion tags'' at each of the transitions in the network indicate the associated feedback that is stored in the memory's evaluation system. Informally, emotion tags can be seen as \emph{remembered rewards} for previous actions. They help the agent to evaluate the result of a  simulation and to translate it into real action. If a clip transition in the simulation leads subsequently to a rewarded action, the state of its tag is set (or confirmed) to $\smiley$, and the transition probability in the next simulation is amplified. Otherwise the tag is set to $\frownie$ and the transition probability is attenuated (or simply not amplified).}
\label{InvasionClipNetwork}
\end{figure}

Suppose that the attacker indicates with the symbols $\Leftarrow$, $\Rightarrow$ that it will move one door to the left, or to the right, respectively. Then, the episodic memory that will be built up by the agent has the graph structure as shown in  Figure \ref{InvasionClipNetwork}.

\subsection{Projective simulation \& learning without composition}
\lb{ssec:no-composition}

As we have mentioned earlier, the interaction of the agent with the environment goes in cycles. In our simple example, the description of the $n$th cycle (or time step) is as follows: First, the agent perceives a percept $s$, which induces the excitation of the percept clip \tc{$s$}. Here we assume that this excitation happens with unit probability, which corresponds to a simple choice for the input coupler function ${\cal I}(c|s)=\delta(c-\mbox{\tc{$s$}})$ introduced in Section~\ref{SectionFormalDefinitions}. The excited percept clip \tc{$s$} then triggers the excitation of action clip \tc{$a$} $\in \{\mbox{\tc{$-$}}, \mbox{\tc{$+$}}\}$ with probability $p^{(n)}(a|s)$. This can happen either in direct sequence, or after some other memory clips have been excited in between, as will be described in the following section. The excitation of an actuator clip \tc{$a$} usually leads to immediate (real) motor action $a$, corresponding to a simple choice for the output coupler ${\cal O}(a|c)=\delta(c-\mbox{\tc{$a$}})$ of Section~\ref{SectionFormalDefinitions}. But we will also consider different scenarios where the translation into motor action may be delayed and depend itself on the emotional tag of the transition \tc{$c$}$\to $\tc{$a$}, resulting from a reward or penalty of that transition in previous cycles. After motor action $a$ has been taken, it will either be rewarded or  not. The result of this evaluation will then be fed back into the state of the episodic memory, leading to an update of the transition probabilities $p^{(n+1)}(a|s)$ for the next cycle and of the emotion state tagged to this transition. This completes the description of the $n$-th cycle.

To provide a complete description of the episodic memory we now need to specify the update rules, i.e.\ how a positive or negative reward ($\Lambda = 1$ or $0$) changes the transition probability between the associated clips. There are many choices possible. In the following, we choose a simple \emph{frequency rule}, somewhat reminiscent of Hebbian learning in neural network theories, but we emphasize that other rules are equally suitable \footnote{A different example is the ``halving rule'' where, after a positive (negative) feedback, the transition probabilities used in the simulation are increased (decreased) by halving the distance to their maximum (minimum) value. This increases both the learning and the adaption speed of the agent substantially. At the same time, there is no notion of a ``long-term'' or ``short-term'' memory, which could be preferential in other scenarios, since the agent learns very fast.}.

We assume that, under positive feedback, the conditional probabilities $p^{(n)}(a|s)$, with $a\in \{-,+\}$, $s\in \{\Leftarrow,\Rightarrow\}$, grow in proportion with the number of previous rewards following the clip transition \tc{$s$} $\to $ \tc{$a$}.
This means that, if, in time step $n$, the agent takes the rewarded action $a$ after having perceived percept $s$, this will increase the probability that, in subsequent time step $n+1$, an excited percept \emph{clip} \tc{$s$} will excite an actuator \emph{clip }\tc{a}. In other words, this will increase the probability that, after perceiving the percept $s$ \emph{next time}, the agent will \emph{simulate the correct action} $a$.
Depending on the details how the simulation is translated into real action, this will typically also increase the probability that the agent executes the rewarded action. Note, however, that the distinction between simulated action and real action is an essential point and will give the agent more flexibility.

Quantitatively, we define the transition probability $p^{(n)}(a|s)$ in terms of a \emph{weight matrix} $h$:
\begin{equation}\label{WeightMatrix}
p^{(n)}(a|s) = \frac{h^{(n)}(s,a)}{h^{(n)}(s)}\, ,
\end{equation}
where $h^{(n)}(s)$ is the marginal
\begin{equation}
h^{(n)}(s) = \sum_{a\in A} h^{(n)}(s,a)\, .
\end{equation}
The weight matrix is, unless otherwise specified, initialized as
\begin{equation}
h^{(1)}(s,a)=1\quad \forall a,s\, ,
\end{equation}
so that the conditional probability distributions $\{p^{(1)}(a|s)\}_{a}$ are uniform for all $s$.

The stepwise evolution of $p^{(n)}(a|s)$, as a function of $n$, is stochastic and may, for a given agent, depend on the entire history of percepts and the actions taken by the agent. Suppose that, in time step $n$, the agent perceives symbol $s^{(n)}$ and then executes action $a^{(n)}$. There are two possible cases which we need to distinguish.\\
\emph{Case (1)}: $\Lambda(s^{(n)},a^{(n)}) = 1$, i.e.\ the agent did the ``right thing'' and the percept-action sequence $(s^{(n)},a^{(n)})$ is rewarded.
In this case, the weight of the $h$ matrix will be increased by unity on the transition \tc{$s$} $\to $\tc{$a$} with $s=s^{(n)}$ and $a=a^{(n)}$, while it stays constant on all other transitions. To model the possibility that the agent can also \emph{forget}, we introduce an overall dissipation factor $\gamma$ $(0\le \gamma \le 1)$ that drives the weights $h^{(n)}(s,a)$ towards the equilibrium (uniform) distribution. Put together we thus have the update rule:
\begin{eqnarray}\label{Rewarded}
h^{(n+1)}(s,a) - h^{(n)}(s,a) =  \delta(s,s^{(n)})\delta(a,a^{(n)}) \\ \nonumber
- \gamma [h^{(n)}(s,a)-1] .
\end{eqnarray}
\emph{Case (2)}: $\Lambda(s^{(n)},a^{(n)}) = 0$, i.e.\ the agent did the ``wrong thing'' and the percept-action sequence $(s^{(n)},a^{(n)})$ is not rewarded.
In this case, all weights of the $h$-matrix are simply decreased:
\begin{eqnarray}\label{Damped}
h^{(n+1)}(s,a) - h^{(n)}(s,a) = - \gamma [h^{(n)}(s,a)-1] .
\end{eqnarray}
The two cases can be combined into a single formula
\begin{eqnarray}\label{RewardedDamped}
h^{(n+1)}(s,a) - h^{(n)}(s,a) =  - \gamma [h^{(n)}(s,a)-1] \nonumber \\
+ \lambda \delta(s,s^{(n)})\delta(a,a^{(n)})
\end{eqnarray}
with $\lambda \equiv \Lambda(s^{(n)},a^{(n)})$, which also generalizes to a situation with values of the reward function $\Lambda$ different from $0$ and $1$.

From the updated weights $h^{(n+1)}(s,a)$, we obtain the transition probabilities (in clip space) for the next cycle,
\begin{equation}
p^{(n+1)}(a|s) = \frac{h^{(n+1)}(s,a)}{\sum_{a} h^{(n+1)}(s,a)} .
\end{equation}

The updating of the weights from $h^{(n)}(s,a)$ to $h^{(n+1)}(s,a)$ at the end of cycle $n$ thus depends on which specific percept-action sequence $(s^{(n)},a^{(n)})$ has actually occurred in cycle $n$. The probability for the latter is given by the joint probability distribution $P^{(n)}(s,a)=P^{(n)}(s) P^{(n)}(a|s)$ for  $(s,a)=(s^{(n)},a^{(n)})$. While $P^{(n)}(s)$ will be given externally (it is controlled by the attacker, for example $P^{(n)}(s)=1/|S|$ for random attacks), the conditional probability $P^{(n)}(a|s)$ will depend on the memory, that is, on the weights $h^{(n)}(s,a)$ and how the simulation is translated into real action.

In the simplest model, the agent has reflection time 1, which corresponds to the following process. Initially the percept $s$ activates the percept clip \tc{$s$}. This excites the actuator clip \tc{$a$} with probability $p^{(n)}(a|s)$. Regardless of whether the action $a$ was previously rewarded or not, \tc{$a$} is coupled out, i.e., it is translated into the action $a$. In other words, any transition that ends up in a clip describing some ``virtual action'', leads to the corresponding {\em real} action. In this case, we obtain
\begin{eqnarray}
P^{(n)}(a|s)=p^{(n)}(a|s) = \frac{h^{(n)}(s,a)}{\sum_{a} h^{(n)}(s,a)} ,
\label{CondProb_R1}
\end{eqnarray}
which complements the update rules of Eqs.~(\ref{Rewarded}) and
(\ref{Damped}), together with Eq.~(\ref{WeightMatrix}).

A slightly more sophisticated model is obtained when the state of the emotion tags ($\smiley$ or $\frownie$), which is set by previous rewards, is used to affirm or inhibit immediate motor action. In this model, the memory is one step further detached from immediate action and the agent has a chance to ``reflect'' upon its action.  To be specific, let us consider a strategy with reflection time  $R$, which corresponds to the following process. As in the previous case, initially the percept $s$ activates the percept clip \tc{$s$}, which activates the actuator clip \tc{$a$} with probability $p^{(n)}(a|s)$.  However, only if the sequence \tc{$s$} $\to $ \tc{$a$} is tagged $\smiley$ (i.e. it was evaluated $\Lambda(s,a)=1$ on the last encounter), the actuator clip \tc{$a$} is ``coupled out'', i.e.\ translated into a real action.
If this is \emph{not} the case (either the transition was not evaluated before or it was evaluated $\frownie$\footnote{If the reward situation changes in a certain point, e.g. due to a change of strategy of the attacker, the tagging of a given clip transition will adapt subsequently.}), the percept clip \tc{$s$} is re-excited, which in turn activates again some actuator clip \tc{$a'$} (where \tc{$a'$} and \tc{$a$} may be the same or different). If the new sequence $(s,a')$ is tagged $\smiley$, \tc{$a'$} triggers real actuator motion $a'$. Otherwise, the process is again repeated. For a model with reflection time $R$, the maximum number of repetitions is $R-1$. At the end of the $R$th round, the simulation must exit from any actuator clip, regardless of its previous evaluations. We are specifically interested in the success probability $P^{(n)}(a^{*}_s|s)$ that the agent chooses a rewarded action $a^{*}_s$ after a given percept $s$  ($\Lambda(s,a^{*}_s)=1$). For reflection time $R$, this is given by
\begin{eqnarray}
P^{(n)}(a^{*}_s|s)=1-\left(1-p^{(n)}(a^{*}_s|s)\right)^R ,
\label{CondProb_RRrewarded}
\end{eqnarray}
which increases with $R$. Clearly, for larger reflection times the memory is used more efficiently.

In our invasion game, the quantity of interest is the blocking efficiency, $r^{(n)}$, which corresponds to the average success probability (averaged over different percepts, i.e. symbols shown by the attacker). After the $n$th round, the blocking efficiency is thus given by
\bea
r^{(n)}= \sum_{s\in S}P^{(n)}(s) P^{(n)}(a^{*}_s|s).
\label{BlockingEfficiency}
\eea

In a similar way one can define the \emph{learning time} $\tau(r_{th})$ for a given strategy as the time it takes on average (over an ensemble of identical agents) until the blocking efficiency reaches a certain threshold value $r_{th}$.

\begin{figure}[tb]
\includegraphics[width=9cm]{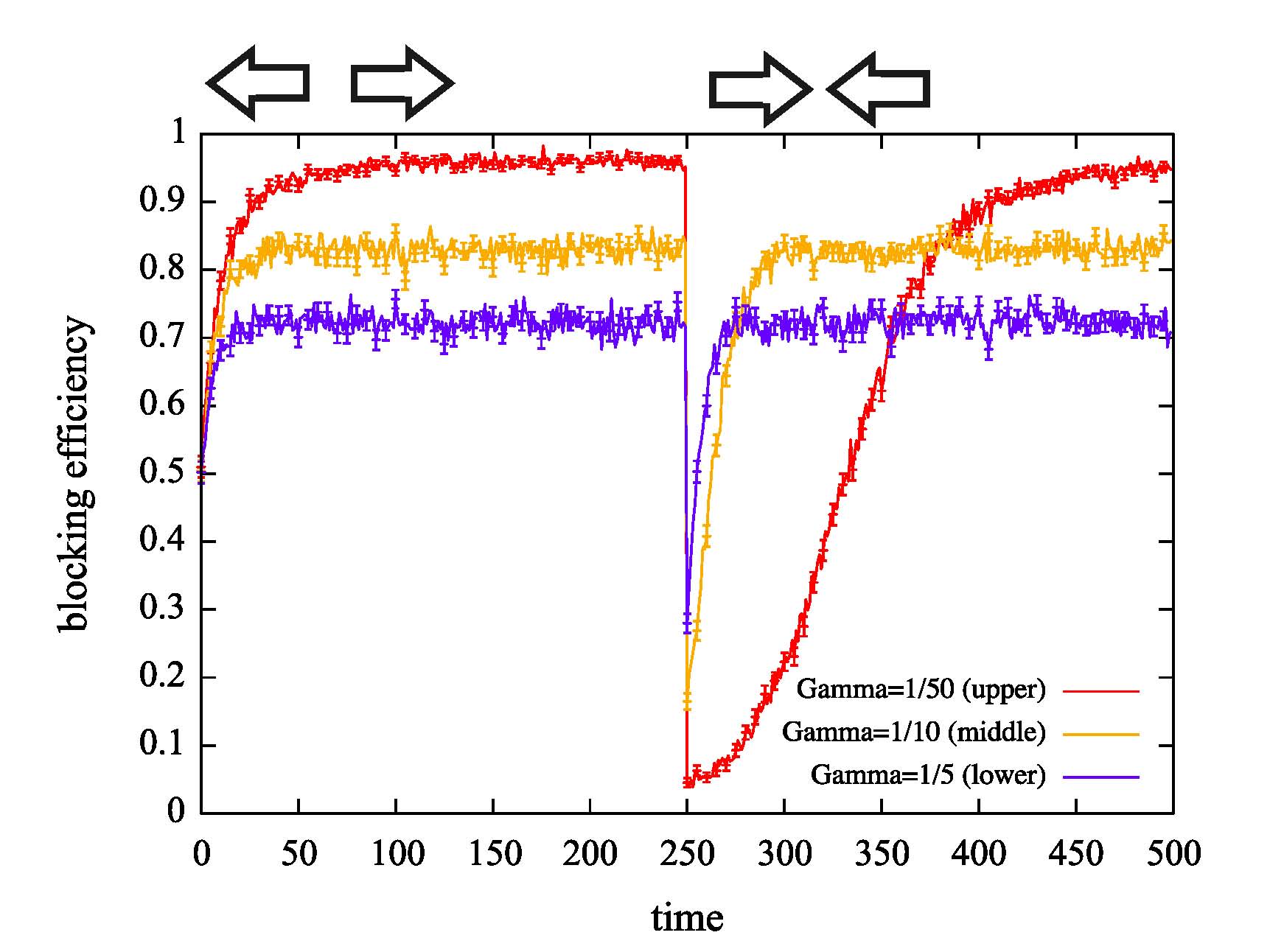}
\caption{Learning curves of the defender agent for different values of the dissipation rate $\gamma$. The blocking efficiency increases with time and approaches its maximum value exponentially fast in the number of cycles. For $\gamma=0$ the blocking efficiency approaches the limiting value 1, i.e. for each shown percept it will choose the right action. For larger values of $\gamma$, the maximum achievable blocking efficiency is reduced, since the agent \emph{forgets} part of what it has learnt. At time step $n=250$, the meaning of symbols is \emph{inverted}, i.e. the symbol $\Rightarrow$ ($\Leftarrow$) now indicates that the attacker is going to move left (right). Since the agent has already built up memory, it needs some time to adapt to the new situation. One can see a trade-off between adaptation speed, one one side, and achievable blocking efficiency, on the other side. Here, we have chosen an unbiased training strategy, $P^{(n)}=1/|S|$. The curves are averages of the learning curves for an ensemble of 1000 agents. Error bars (indicating 1 standard deviation over the sample mean) are shown on every fifth data point not to clutter the diagram, which also applies to the error bars in subsequent Figures.
}
\label{FIG_SimpleLearning}
\end{figure}

In the following, we show numeric results for different agent specifications. Let us start with agents with reflection time $R=1$.
In Figure \ref{FIG_SimpleLearning}, we plot the learning curves for different values of the dissipation rate $\gamma$ (forgetfulness). One can see that the blocking efficiency increases with time and approaches its maximum value typically exponentially fast in the number of cycles.
For small values of $\gamma $ it approaches the limiting value 1, i.e.\ the agent will choose the right action for every shown percept. For increasing values of $\gamma$, we see that the maximum achievable blocking efficiency is reduced, since the agent keeps forgetting part of what it has learnt. At time step $n=250$, the attacker suddenly changes the meaning of symbols: $\Rightarrow$ ($\Leftarrow$) now indicates that the attacker is going to move left (right). Since the agent has already built up memory, it needs some time to adapt to the new situation. Here, one can see that forgetfulness can also have a positive effect. For weak dissipation, the agent needs longer to \emph{unlearn}, i.e.\ to dissipate its memory and adapt to the new situation. Thus there is a trade-off between adaptation speed, on one side, and achievable blocking efficiency, on the other side. Depending on whether learning speed or achievable efficiency is more important, one will choose the agent specification accordingly. Note that for random action, which is obtained by setting $\lambda = 0$ in (\ref{RewardedDamped}), the average blocking is $0.5$ (not shown in Figure \ref{FIG_SimpleLearning}).

\begin{figure}[tb]
\includegraphics[width=9cm]{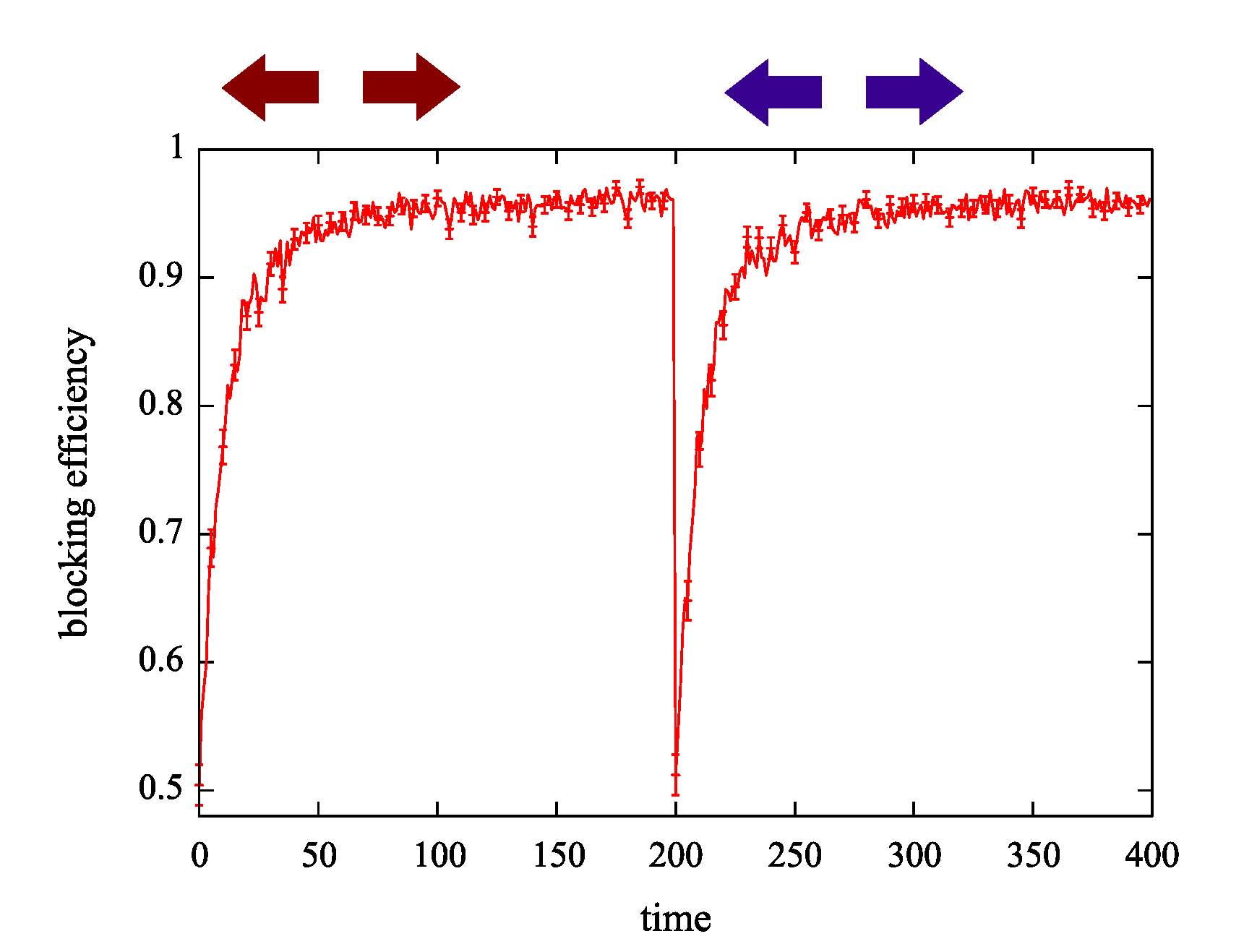}
\caption{Learning curve for enlarged percept space, with color as an additional percept category. In the first period, the symbols seen by the agent have the same color (e.g. red), while at time step $n=200$ the color of the symbols suddenly changes (e.g. blue), and the agent has to learn the meaning of the symbol with the new color. Unlike Figure \ref{FIG_SimpleLearning}, there is no inversion of strategies, and thus no increased adaptation time. The agent simple has not seen symbols with the new color before, and thus has to learn them from scratch. Ensemble average over 1000 runs with error bars indicating one standard deviation.}
\label{FIG_TwoColourLearning}
\end{figure}

Note that the existence of an adaptation period in Figure \ref{FIG_SimpleLearning} (after time step $n=250$) relates to the fact that symbols which the agent \emph{had already learnt}, suddenly invert their meaning in terms of the reward function. So the learnt behavior will, with high probability, lead to unrewarded actions. A different situation is of course given, if the agent is confronted with a \emph{new} symbol that it had not perceived before. In Figure \ref{FIG_TwoColourLearning}, we have enlarged the percept space and introduced color as an additional percept category. In terms of the invasion game, this means that the attacker can announce its next move by using symbols of different shapes \emph{and} colors. In the first period, the symbols seen by the agent have a specific color (red), while at $n=250$ the color suddenly changes (blue), and the agent has to learn the meaning of the symbols with the new color. Note that, unlike Figure \ref{FIG_SimpleLearning}, there is now no inversion of strategies, and thus no increased adaptation time. The agent simply has never seen blue symbols before, and has to learn their meaning from scratch \footnote{A similar plot as in Figure~\ref{FIG_TwoColourLearning} is obtained if the attacker uses symbols of a single color only, but shows in the first period (up to $n=249$) only the symbol $\Rightarrow$ and in the second period (from $n=250$ on) only the symbol $\Leftarrow$.}.

The network behind Figure \ref{FIG_TwoColourLearning} is the same as in Figure~ \ref{InvasionClipNetwork}, with the same update rules, but with an extended percept space (four symbols) and four rewarded transitions. The agent does not make use of the ``similarity'' between symbols with the same shape but with different colors.
This will change in the next section, when we introduce the idea of composition as another feature of projective simulation, which will allow us to realize an elementary example of associate learning.

\begin{figure}[htb]
\includegraphics[width=9cm]{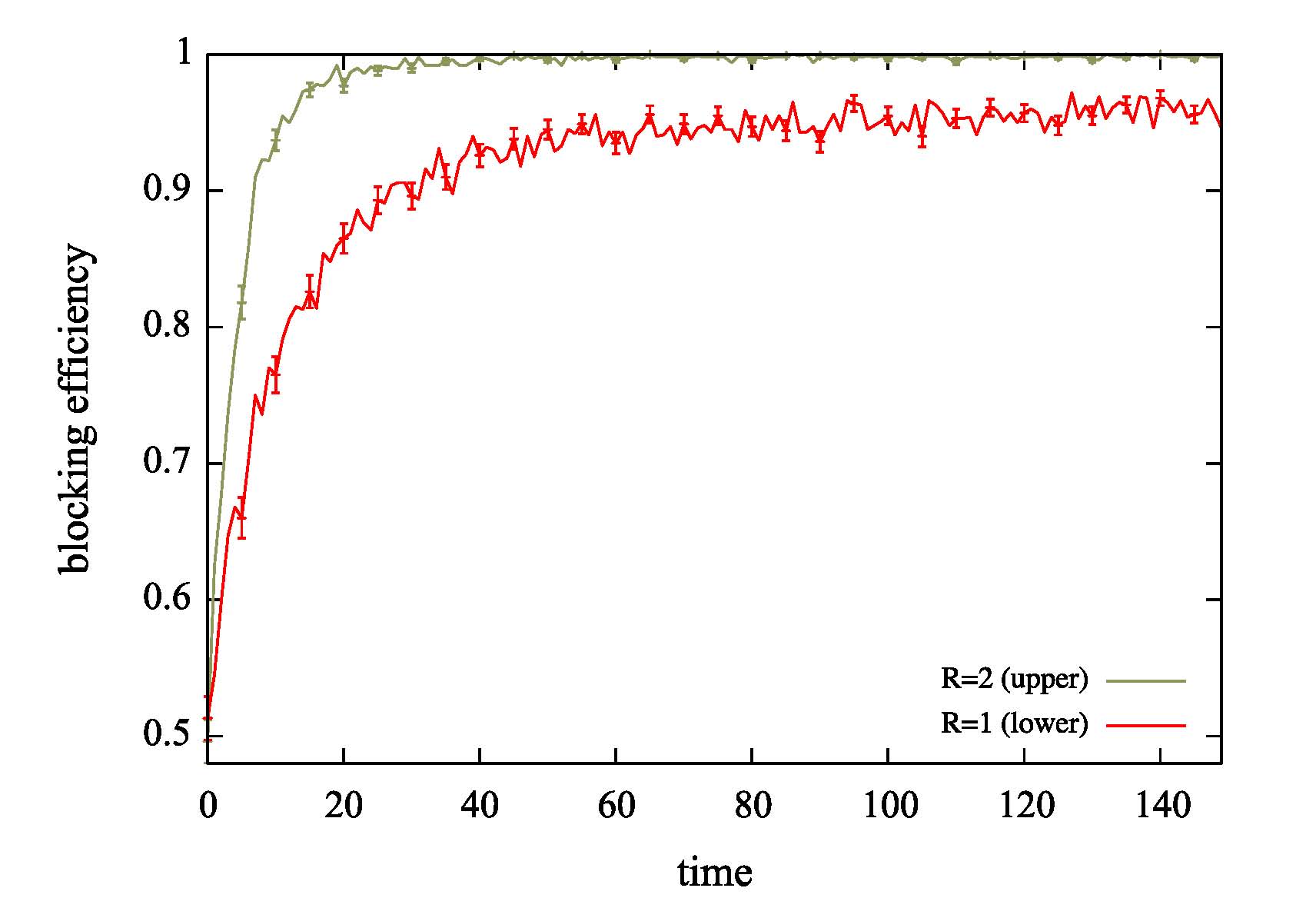}
\caption{Performance of agents with different values of the reflection time: $R=1$
(lower curve) and $R=2$ (upper curve). One can see that a large value of the reflection time leads to an increased learning speed. The dissipation rate (which is a measure of forgetfulness of the agent) is in both cases $\gamma = 1/50$. Ensemble average over 1000 runs with error bars indicating one standard deviation.}
\label{FIG_ReflectionTimeLearning}
\end{figure}

Let us now come back to the notion of reflection. In Figure \ref{FIG_ReflectionTimeLearning}, we compare the performance of agents with different values of the reflection time $R$. (Here we consider again training with symbols of a single color.) One can see that larger values of the reflection time lead to an increased learning speed. The reason is that during the simulation virtual percept-action sequences are recalled together with the associated emotion tags (i.e.\ remembered rewards).
If the associated tag does not indicate a previous reward of the simulated transition, the coupling-out of the actuator into motor action is suppressed and the simulation goes back to the initial clip. In this sense, the agent can ``reflect upon'' the right action and its (empirically likely) consequences by means of an iterated simulation, and is thus more likely to find the right actuator move before real action takes place \footnote{Of course, one could summarize the entire result of the simulation into an effective update rule for the agent's policy $P^{(n)}(a|s)$, not referring to the internal rules of episodic memory recall. In this sense, we are just describing a special type of learning agent. The point is, however, that we want to illustrate the idea of projective simulation and its  flexibility in the specific context of a learning agent.}.

The possibility of reflection can thus significantly increase the speed of learning, at least as long the total time for the simulation does not become too long and starts competing with other, externally given time scales, such as frequency of attacks.

\medskip

\begin{figure}[htb]
\hspace*{-0.5cm}\includegraphics[width=9.5cm]{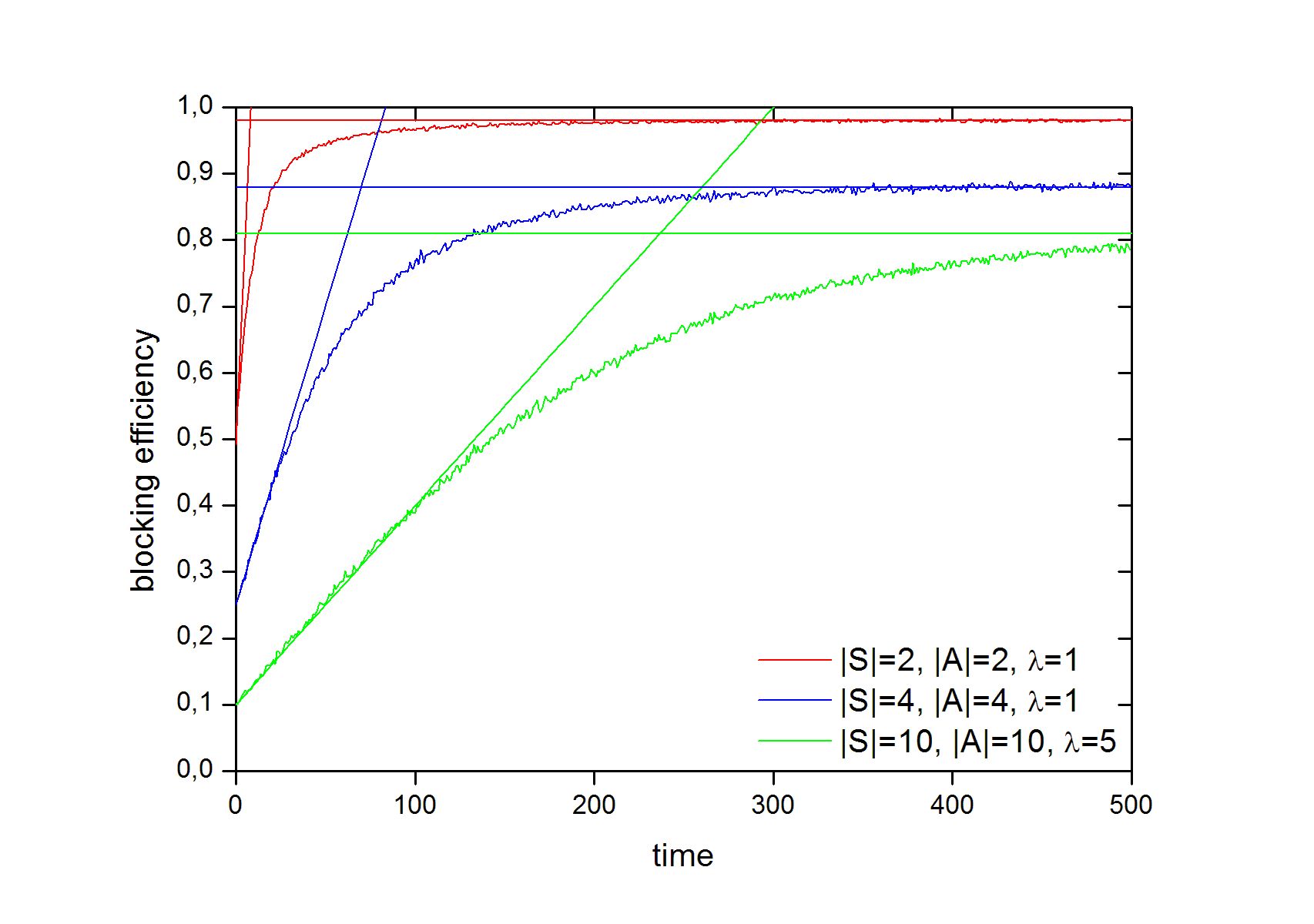}
\caption{Initial growth and asymptotic value of average blocking efficiency for different sizes of percept ($|S|$) and actuator ($|A|$) space, and reward parameter $\lambda$.
The learning curves are obtained from a numerical average over an ensemble of 10000 runs with random percept stimulation ($\gamma = 0.01$). Error bars (not shown) are of the
order of the fluctuations in the learning curves). The analytic lines are obtained from (\ref{MeanFieldRecursion}), see main text.}
\label{FIG_Julian1}
\end{figure}

Within an approximate analytical treatment, one can give a closed recursion relation for the mean entries of the h-matrix.
Consider the general case of $|S|$ different percepts and $|A|$ different actions, where for each percept there is a single rewarded action. For simplicity, let us assume a regular training scenario, $P^{(n)}(s)=\delta(s-n \bmod |S|)$ such that, within a subsequence of $|S|$ cycles,
each percept is excited exactly once and in the same order.
For such a scenario, one can derive from (\ref{RewardedDamped}) a recursion relation of the form
\begin{eqnarray}
\bar{h}^{(n+|S|)}(s,a) - 1  & \simeq &   (1-\gamma)^{|S|}\left(\bar{h}^{(n)}(s,a)-1\right) \nonumber \\
 & + & (1-\gamma)^{|S|-1} \lambda  \frac{\bar{h}(s,a)}{\sum_{a'\in A} \bar{h}(s,a')}
\label{MeanFieldRecursion}
\end{eqnarray}
for rewarded transitions, and a similar expression, without the gain term (i.e.\ $\lambda = 0$), for the unrewarded transitions.
Here, $\bar{h}^{(n)}(s,a)$ denotes the averaged weight for a rewarded transition \tc{$s$} $\to $ \tc{$a$}, taken over an ensemble
of runs. Equation (\ref{MeanFieldRecursion}) is not exact and in general
contains an overestimation of the gain term, but for small values of $\gamma$ it gives a rather good approximation to the
numerical results \cite{ToBePublished}.
The steady-state condition reads $\bar{h}^{(n+|S|)}(s,a) = \bar{h}^{(n)}(s,a) \equiv \bar{h}(s,a)$, whereby
$\bar{h}(s,a')=1$ for all unrewarded transitions. This leads to quadratic equations of the form
\begin{eqnarray}
\bar{h}(s,a) - 1  & = &   (1-\gamma)^{|S|}\left(\bar{h}(s,a)-1\right) \nonumber \\
& + & (1-\gamma)^{|S|-1} \lambda \frac{\bar{h}(s,a)}{\bar{h}(s,a)+|A|-1} ,
\label{MeanFieldRecursion2}
\end{eqnarray}
that can be solved analytically, providing an approximate value for the steady-state blocking efficiency
$\bar r \simeq \frac{\bar{h}}{\bar{h}+|A|-1}$ shown in Figure~\ref{FIG_Julian1}. (For $\lambda = 0$, one obtains from (\ref{MeanFieldRecursion2})
the trivial steady-state value $\bar{h}(s,a)=1$, recovering the value $\bar r=0.5$ for random action). Similarly, based on (\ref{MeanFieldRecursion}),
one can derive an approximate analytic expression for the initial slope of the learning curve
\begin{equation}
\left( \frac{\Delta \bar r}{\Delta n} \right)_{n=1}  =  \frac{\lambda(1-\gamma)^{|S|-1}(|A|-1)}{|A|^3(|S|-1+|S|/2)} .
\end{equation}
In Figure~\ref{FIG_Julian1}, we plot the learning curves (evolution of the average blocking efficiency) together with the analytic approximations, for different values of $|S|, |A|$, and $\lambda$.

\begin{figure}[htb]
\includegraphics[width=9cm]{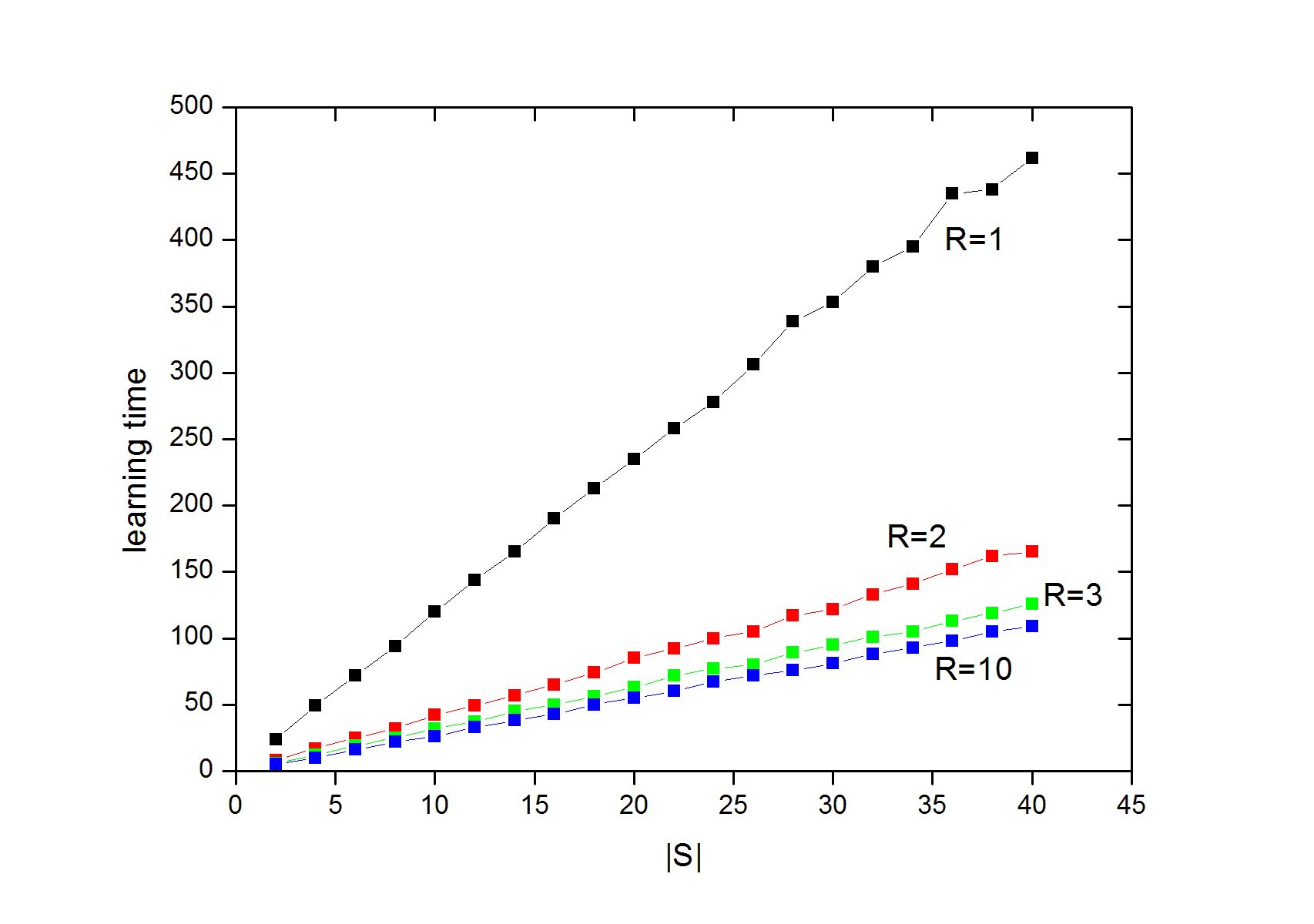}
\caption{Learning time $\tau_{0.9}$ as a function of $|S|$ for different values of the reflection parameter $R$. We observe a linear dependence of $\tau_{0.9}$ on $|S|$ with a slope determined by $R$. Ensemble average over 10000 runs, $\gamma = 0$.}
\label{FIG_LinearScaling}
\end{figure}

We next investigate the performance of the agent for more complex environment in order to illustrate the scalability of our model.
In the invasion game, a natural scaling parameter is given by the size $|S|$ of the percept space
(number of doors through which attacker can invade) and/or the size $|A|$ of the actuator space. As a figure of merit, we have looked at the learning time $\tau=\tau_{0.9}$, which
we define as the time the agent needs to achieve a certain blocking efficiency (for which we choose $90\%$ of the maximum achievable value).
We find that learning time increases linearly in both $|S|$ and $|A|$, (i.e.\ quadratically in $N$, if we set $N=|A|=|S|$). The same scaling can be observed if we apply
standard learning algorithms like Q-learning or AHC \cite{SuttonBarto98} to the invasion game \cite{ToBePublished}.
In Figure \ref{FIG_LinearScaling}, the scaling of the learning time is shown for different values of $R$. Besides the linear scaling with $|S|$, it can be seen how reflections in clip space, as part of the simulation, speed up the learning process.

A more detailed discussion of the analytic results, together with a comparison of PS with other models of
reinforcement learning, will be presented in \cite{ToBePublished} (see also Section \ref{SectionLiterature}).

\subsection{Projective simulation \& learning with composition I}
\lb{ssec:composition}

The possibility of multiple reflections, as discussed in the previous subsection
(Figure~\ref{FIG_ReflectionTimeLearning}), illustrates an advantage of having a simulation platform where previous experience can be reinvoked and evaluated before real action is taken.

The episodic memory described in Figure~\ref{InvasionClipNetwork} was of course a quite elementary and special instance of the general scheme of Figure~\ref{ClipNetwork}. We have assumed that the activation of a percept clip is immediately followed by the activation of an actuator clip, simulating a simple percept-action sequence. This can obviously be generalized along various directions. In the following, we shall discuss one generalization, where the excitation of a percept clip may be followed by a sequence of jumps to other, intermediate clips, before it ends up in an actuator clip. These intermediate clips  may correspond to similar, previously encountered percepts, realizing some sort of associative memory, but they may also describe clips that are spontaneously created and entirely fictitious (see Section~\ref{ssec:composition}).

Such a scenario, which generalizes the situation of Figure~\ref{InvasionClipNetwork}, can be summarized by the following rules.

\begin{figure}[tb]
\centering
\psfrag{A}{\small{Direct transitions}}
\psfrag{B}{\small{Composition}}
\psfrag{M}{\small{Memorized-- or}}
\psfrag{m}{\small{fictitious clips}}
\psfrag{C}{\small{Actuator clips}}
\psfrag{c}{\small{Percept clip}}
\psfrag{2}{\small{Clip$'$}}
\psfrag{3}{\small{Clip$''$}}
\psfrag{4}{\small{Clip$'''$}}
\psfrag{5}{\small{Clip$^{iv}$}}
\psfrag{6}{\small{Clip$^v$}}
\psfrag{7}{\small{Clip$^{vi}$}}
\includegraphics[width=0.9\columnwidth]{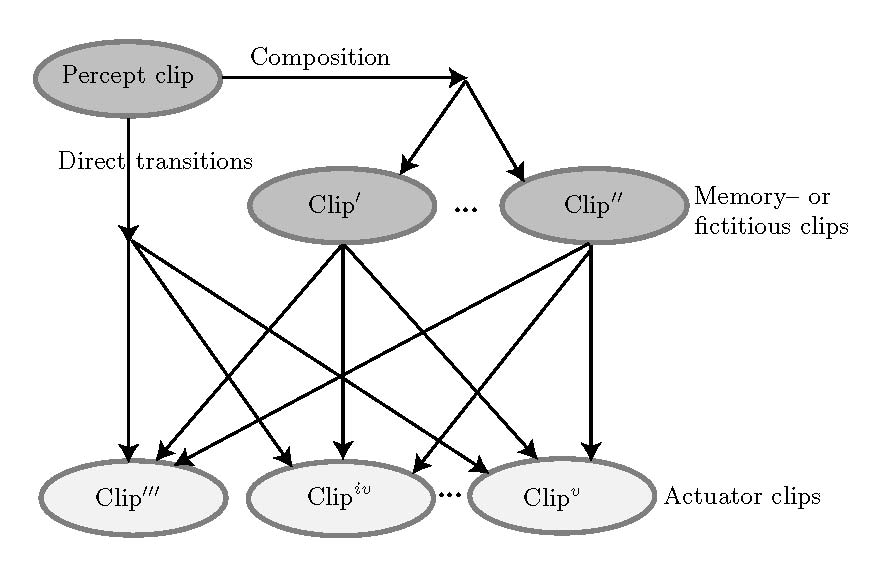}
\caption{Projective simulation with composition with deliberation length $D=0,1$. Dark gray ovals indicate percept clips and light dark ovals indicate actuator clips. Initially the percept clip is excited. This may directly excite some actuator clip (``Direct transitions''), or some other memory clip or fictitious clip (``Composition''). In the latter case, the memory (or fictitious) clip in its turn excites an actuator clip.}
\label{Fig_DirectvsComposition}
\end{figure}

\begin{enumerate}
\item
Every percept $s$ triggers a sequence of memory clips
$\Gamma$ = (\tc{$s$},\tc{$s^{1}$},\dots, \tc{$s^{D}$}, \tc{$a$}), starting with \tc{$s$} and ending with some actuator clip \tc{$a$} \footnote{The case in which an actuator clip is not only excited at the end of a simulation, but also within, can be considered as well. For simplicity, we will here focus our attention to described sequences $\Gamma$. However, multiple excitations of actuator clips within a simulation do occur in the context of multiple reflections.}. The number $D$ denotes the \emph{deliberation length} of the sequence. The case $D=0$ corresponds, per definition, to the direct sequence $\Gamma$ = (\tc{$s$},\tc{$a$}).

This is illustrated schematically in Figure \ref{Fig_DirectvsComposition}, where we show an example of an episodic memory architecture with sequences of deliberation length $D=0$ and $D=1$ is shown. Here, after excitation of the percept clip, the agent may either excite an actuator clip directly, or first excite some other intermediate clip which, in its turn, activates an actuator clip. We shall sometimes refer to the former sequence as ``direct'', and to the latter as ``compositional''.

\item
If $(s,a)$ corresponds to a rewarded percept-action pair (i.e.\ it was rewarded in a recent cycle and the corresponding emotion tag is set to $\smiley$) \footnote{Note that there is a certain freedom as to which part of the sequence the tag should be associated. A simplest choice, which we follow here, is that the tag refers only to the states of the initial and the final clip.}, then the simulation is left and the actuator clip \tc{$a$} is translated into real action $a$. Otherwise, a new (random) sequence $\Gamma'$ = (\tc{$s$},\tc{$s^{1'}$},\dots, \tc{$s^{D'}$}, \tc{$a'$}) is generated, starting with the same percept clip \tc{$s$} but ending possibly with a different actuator clip \tc{$a'$}. The (maximum) number of fictitious clip sequences that may occur before real action is taken is given by the reflection time $R$.

\item
The probability for a transition from clip \tc{$c$} to clip \tc{$c'$} is determined by the weights $h^{(n)}(c,c')$ of the edges of a directed graph
\cite{GodsilRoyle01} connecting the corresponding clips:
\begin{equation}
p^{(n)}(c'|c)= \frac{h^{(n)}(c,c')}{\sum_{c''} h^{(n)}(c,c'')}
\end{equation}
where the sum in the denominator runs over all clips \tc{$c''$} that are connected with \tc{$c$} by an \emph{outgoing} edge (i.e. an edge directed from \tc{$c$} to \tc{$c''$}).

\item
After the simulation in cycle $n$ is concluded, some action will be taken which we denote by $a^{(n)}$. If the action $a^{(n)}$ is \emph{rewarded} (i.e.\ $\Lambda(s^{(n)},a^{(n)})=1$), then the weights of all transitions that
occurred in the preceding simulation will be enhanced:
\newline
(i) The weights of transitions \tc{$c$} $\to $ \tc{$c'$} that appear in the simulated sequence $\Gamma$=(\tc{$s$}, \dots, \tc{$c$}, \tc{$c'$}, \dots,  \tc{$a$}) with $s=s^{(n)}$ and $a=a^{(n)}$ increase by the amount
\begin{eqnarray}
\Delta_{+}h^{(n)}(s^{i},s^{i+1}) &=& K \quad \mbox{for} \quad i=1,...,D-1 , \nonumber\\
\Delta_{+}h^{(n)}(s,s^{1}) &=& \Delta_{+}h^{(n)}(s^{D},a) = 1 .
\label{WeightUpdateRewardedComp}
\end{eqnarray}
\newline
(ii) In addition, the weight of the direct transition \tc{$s$} $\to $ \tc{$a$} will also be increased by unity
\begin{equation}
\Delta_{+}h^{(n)}(s,a)= 1 .
\label{WeightUpdateRewardedDirect}
\end{equation}
The parameter $K$ thereby quantifies the growth rate of ``associative'' (or compositional) connections relative to the direct connections.
\newline
(iii) Furthermore, the weights of \emph{all} transitions in the clip network, including those which were not involved in the preceding simulation,
will be decreased according to the rule
\begin{equation}
\Delta_{-}h^{(n)}(c,c')=-\gamma\left( h^{(n)}(c,c') - h_{0}(c,c')\right) ,
\end{equation}
which describes \emph{damping} towards a stationary value
\begin{equation}
h_{0}(c,c') = { 1, \quad  \mbox{if} \; c\in S \; \mbox{and} \; c'\in A \brace K, \quad \mbox{if} \; c\in S \; \mbox{and} \; c'\in S \hfill} ,
\end{equation}
which distinguishes again direct connections from compositional connections, as illustrated in Figure~\ref{Fig_DirectvsComposition}.
If the chosen action $a^{(n)}$ at the end of cycle $n$ is \emph{not rewarded}, then no weights are enhanced and only rule (iii) applies.

\item
Concerning the initialization of the weights, various possibilities exist. Weights that are initialized to unity describe a sort of ``innate'' or \emph{a priori} connections between a set of basic percepts and actuators. Other weights may initially be set to zero, for example on connections to more complex percepts, for which there are no innate action patterns available.  A simple rule that allows the connectivity of the memory (graph of the clip network) to \emph{grow} through new perceptual input, is the following:

If a percept clip is activated for the first time, all \emph{incoming} connections to that clip are ``activated'' together with it, meaning that their weights are initialized to a finite value (which we also set to $K$ in the following) \footnote{In order that a percept can be perceived at all, it must already have a (potential) representation in memory space. The architecture of memory reflects what is syntactically possible, i.e.\ what can be perceived a priori. We speak of the ``activation'' of a percept clip once it is hit by a real (external) stimulus for the first time. The additional quality of a percept that has already been stimulated is that, in memory space, the corresponding percept clip is not isolated but can be reached from other percepts.}. This enables the accessibility of that clip from other clips.
\end{enumerate}

To illustrate the workings of compositional memory, let us revisit the situation of Figure \ref{FIG_TwoColourLearning}, where the percept space $S=S_1\times S_2$ comprises both the categories of shape, $s_1\in S_1$, and color, $s_2\in S_2$ (the color of the shape), while the actuator space $A$ and the emotion space $E$ contain the same elements as before. This is a variant of the invasion game, where the attacker can announce its next move using symbols of different shapes \emph{and} colors. The network of clips behind the learning curves presented in Figure \ref{FIG_TwoColourLearning} was simply a duplicated version of the graph in Figure \ref{InvasionClipNetwork}, with identical subgraphs for the two sets of percepts of the same color.

\begin{figure}[tb]
\hspace*{-12pt}
\includegraphics[width=9.5cm]{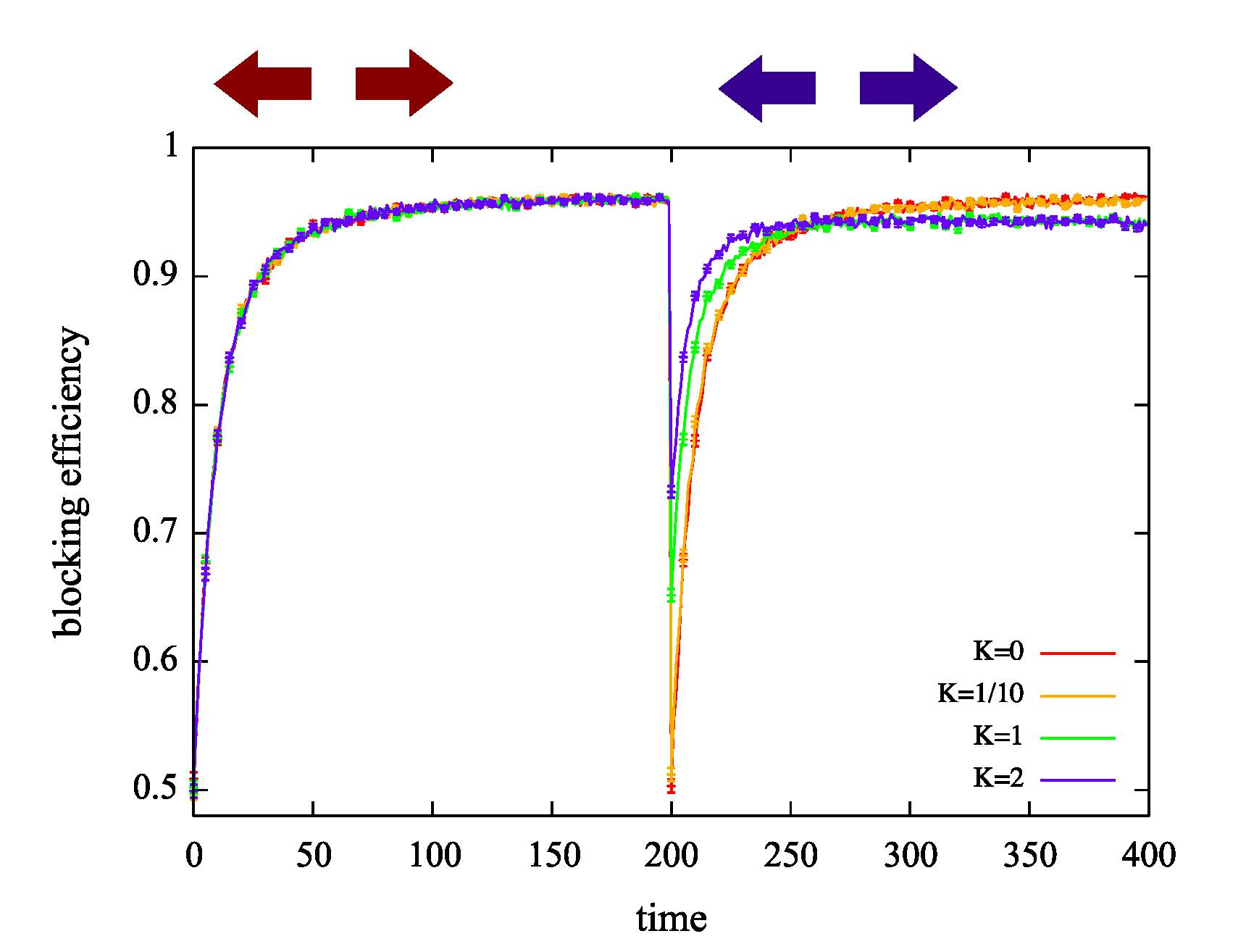}
\caption{Associative learning through projective simulation. After first training the agent with symbols of one color (red), at time step $n=200$ the attacker starts to use a different color (blue). In comparison with Figure \ref{FIG_TwoColourLearning}, now the agent learns faster. This situation resembles a form of ``associative learning'', when the agent ``recognizes'' a similarity between the percepts of different colors, but identical shapes. The effect can be much enhanced if one allows for reflection times $R>1$.  The memory that gives rise to these learning curves is depicted in Figure \ref{FIG_CompAssocMem}. Ensemble average over 10000 agents.}
\label{FIG_CompositionAssociation}
\end{figure}

In contrast, in Figure \ref{FIG_CompositionAssociation}, we see the learning curves for the same game but with a slightly modified memory architecture. After having trained the agent with symbols of one color (red), at time step $n=200$ the attacker starts using a different color (blue). In comparison with Figure \ref{FIG_TwoColourLearning}, now the agent learns faster, and the speed of learning increases with the strength of the parameter $K$. This situation resembles a form of ``associative learning'', where the agent ``recognizes'' a similarity between the percepts of different colors (but identical shapes).

\begin{figure}[htb]
\includegraphics[width=8.5cm]{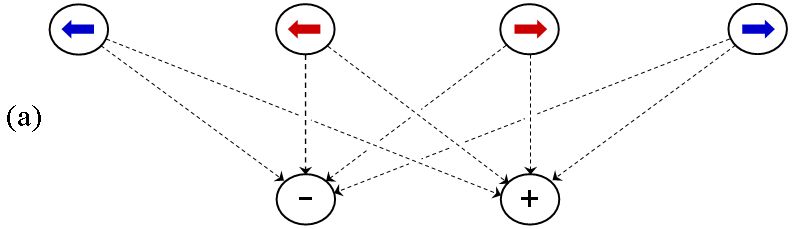} \\
\vspace*{20pt}
\includegraphics[width=8.5cm]{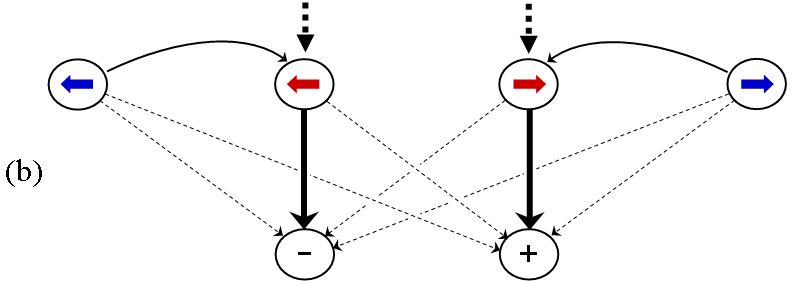} \\
\vspace*{20pt}
\includegraphics[width=8.5cm]{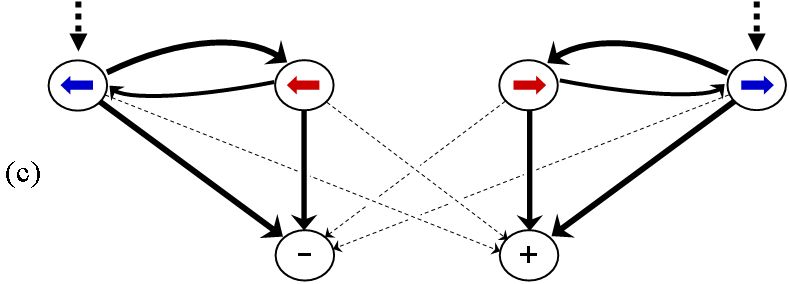}
\caption{Effect of associative learning on the state of the episodic memory at different times. The thickness of the lines indicate the transition probabilities between different clips. (a) Initial network, before any percept has affected the agent, (b) State of the network after the agent has been trained (dotted arrows) with symbols of one color (red). (c) When the agent is presented with symbols of a different color (blue), the established links will direct the simulation process (probabilistically) to the previously ``trained'' region with well-developed links. This realizes a sort of associative memory.}
\label{FIG_CompAssocMem}
\end{figure}

The structure of the memory that gives rise to these learning curves is sketched in Figure~\ref{FIG_CompAssocMem}, which corresponds to a duplicated network described before, albeit with additional links between percepts of equal shape but different color.
In Figure~\ref{FIG_CompAssocMem}, we see the effect of learning on the state of the network at different times. Initially, before any stimulus/percept has affected the agent, the network looks as in Figure~\ref{FIG_CompAssocMem}(a), with innate connections of unit weight between all possible percepts and actuators, respectively.
Figure~\ref{FIG_CompAssocMem}(b) shows the state of the network after the agent has been trained (indicated by the dotted arrows) with symbols of one color (red). We see that the weights for rewarded transitions have grown substantially such
that the presentation of a red symbol will lead to the rewarded actuator move with high probability. Moreover, the activation of the red-percept clips has initialized the incoming connections from similar percept clips with a different (blue) color. In this example, the weights are initialized with the value $K$. This initialization has, at this stage, no effect on the learning performance for symbols with a red color. However, when the agent is presented with symbols of a different color, the established links will direct the simulation process (probabilistically) to a ``trained'' region with well-developed links. This realizes a sort of associative memory (Figure~\ref{FIG_CompAssocMem}(c)). In the philosophy of projective simulation, association is a special instance of a compositional process, namely a random walk in clip space where similar clips can call each other with certain probabilities
\footnote{The network in Figure~\ref{FIG_CompAssocMem}(b) can be seen as a special instance of the network of Figure~\ref{Fig_DirectvsComposition} where the intermediate memory clips correspond to the previously trained (red) percepts, and the percept clip takes the role of the yet-to-be-trained (blue) percept.}.

Note that, in case of the associative learning, only the \emph{incoming} links (i.e. transitions) to that percept are activated together with it, thereby making its subsequent links potentially available to similar new percepts. A network where also outgoing links are activated performs typically worse, in particular when the size of the percept space (number of colors) grows. In that case, even when a single percept is trained, the agent has to explore all similar percepts together with it, which may lead to a significant slowing down of the learning speed.

\begin{figure}[tb]
\includegraphics[width=9cm]{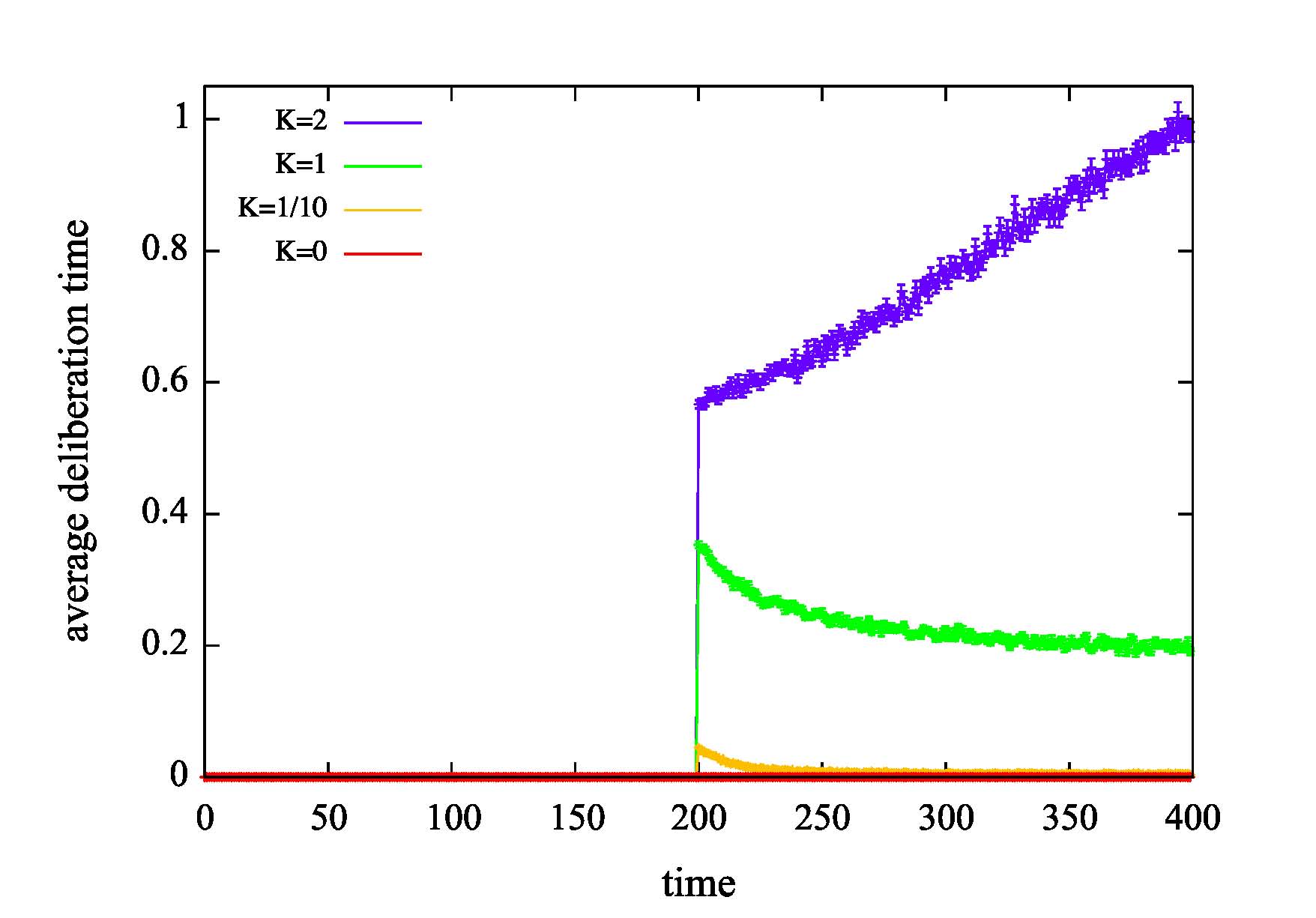}
\caption{Average deliberation time, i.e. the average time how long the simulation stays in compositional memory. A deliberation time that is too long will, in this example, have a negative effect on the learning fidelity as it will
also have an increased access to other, wrong channels. Dissipation rate $\gamma=1/50$; ensemble average over 10000 agents.}
\label{FIG_DwellingTime}
\end{figure}

In Figure~\ref{FIG_DwellingTime}, we discuss further aspects of associative learning that follow from the rules of the projective simulation. We saw in
Figure~\ref{FIG_CompositionAssociation} that the learning speed increases with the parameter $K$, which describes the relative rate at which the weights of the compositional connections grow relative to the direct connections. However, too large values of $K$ can also have a counterproductive effect, as the agent spends an increasing fraction of time with the simulation  before it takes real action. In fact, it can almost get ``lost'' in a loop-like scenario where it jumps back and forth between virtual percept clips for a long time. In Figure~\ref{FIG_DwellingTime}, we plot the average deliberation time, i.e. the average time for which the simulation stays in compositional memory. The scenario is the same as in Figure~\ref{FIG_CompositionAssociation}. After the change of color of the symbols, the agent will learn by building up new transitions in the network, but this learning will be assisted by using the pre-established transitions of the previous training period (Figure~\ref{FIG_CompAssocMem}(c)), which will increase the deliberation time. For $K\le 1$ the deliberation time is maximal right after the change of colors, and decreases again as the agent is developing direct connections from the percept clips to the rewarded actuator clips. For $K=2$, however, the deliberation time continues to grow with the number of cycles, until it settles at some value around 1.4 (not shown). For larger values of $K$, the asymptotic average deliberation time can be significantly larger. In the network of Figure~\ref{FIG_CompAssocMem}(c) the latter situation means that the simulation can get lost in a loop by jumping back and forth between similar (red and blue) clips. While in the simple example of Figure~\ref{FIG_CompAssocMem}(c) this may be avoided by certain \emph{ad hoc} modifications of the update rule, it is a generic feature that will persist in more complex networks.

\begin{figure}[tb]
\begin{center}
\begin{minipage}{9cm}
\includegraphics[width=9cm]{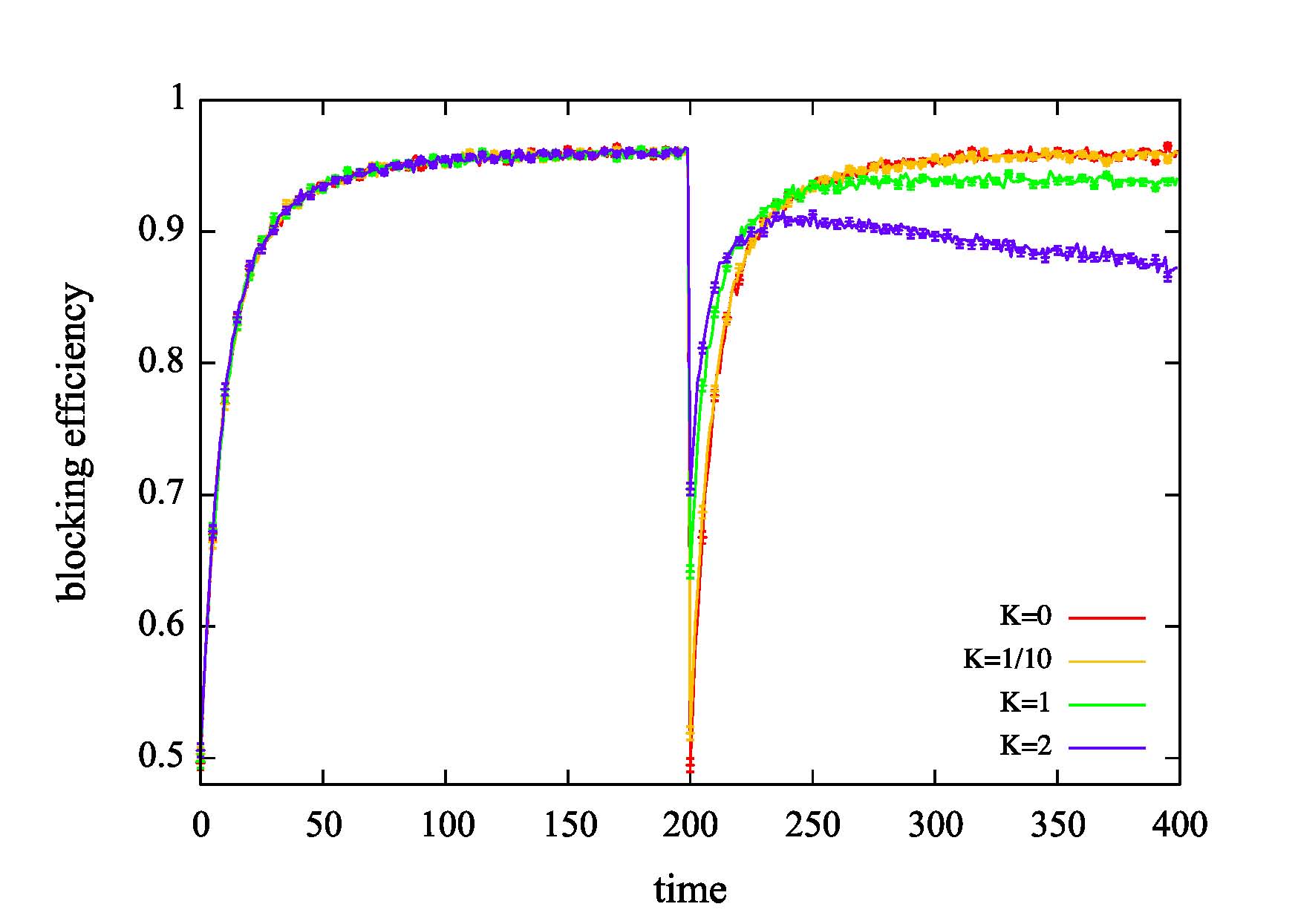}
\end{minipage}
\end{center}
\caption{(a) Learning curve for different values of the associativity parameter $K$ if the agent, by external constraints, has only a finite time available to produce an action. If the simulation takes longer than $D_\texttt{max}$, the agent will not be rewarded. In such a case, the asymptotic performance of the learning drops dramatically for large values of $K$. An ensemble average over 10000 games is shown.}
\label{FIG_MaxDwellingTime}
\end{figure}

A deliberation (i.e. simulation) time that is too long will, in this example,
eventually have a negative effect on the achievable blocking efficiency, as can be seen from the long-time limit of the learning curves in Figure~\ref{FIG_CompositionAssociation}. A slight decrease of the asymptotic blocking efficiency for larger values of $K$ occurs because, by association, the simulation will also gain access to other unrewarded transitions inside the network \footnote{In the specific case of Figure~\ref{FIG_CompAssocMem}(c), it is the transition from the percept clip \tc{{\color{red}$\Leftarrow$}} (red) and the actuator clip \tc{$-$}. Details depend on the relative weights of the outgoing transitions leaving the loop.}. The potentially negative effect of high values of $K$ gets more pronounced if the agent, by external constraints, only has a finite time available to produce an action. In our example of the invasion game, this could be the time it takes for the attacker to move from one door to the next.
This introduces a maximum deliberation time $D_\texttt{max}$ to our scheme. If the simulation takes longer than $D_\texttt{max}$, the agent arrives too late at the door even if it chose the right one, and will consequently not be rewarded. In such a case, the asymptotic performance of the learning for large values of $K$ drops significantly, as can be seen in Figure~\ref{FIG_MaxDwellingTime} for $D_\texttt{max}=2$. For short times, when the strengths of the transitions have not yet grown too large, the simulation still benefits from the association effect where, after jumping from a percept clip \tc{{\color{red}$\Leftarrow$}} (red) to percept clip \tc{{\color{blue}$\Leftarrow$}} (blue), there will be a strong transition to an actuator. For longer times however, the weights on the compositional links have grown so strongly that they will also dominate over the direct links from percept clips to actuator clips. In summary, while compositional memory can help, too large values of $K$ can be counterproductive, as the agent will most of the time be ``busy with itself''.

\begin{figure}[tb]
\includegraphics[width=9cm]{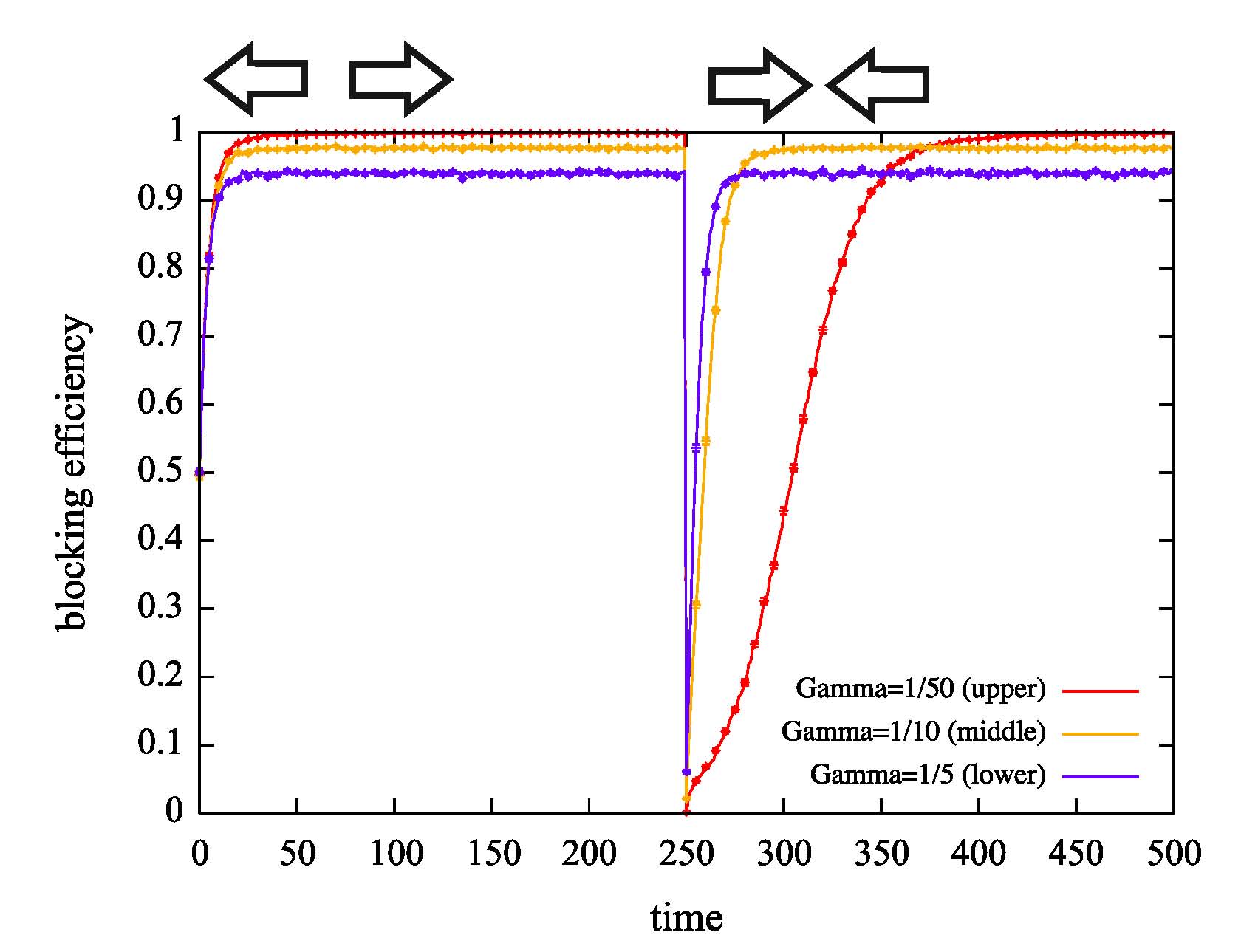}
\caption{To obtain an agent with both high flexibility to adapt to new attack strategies, and with a high blocking efficiency, one can combine a finite dissipation rate $\gamma$ (flexibility) with an increased reflection time $R=2$ (efficiency). The plots should be compared with Figure~\ref{FIG_SimpleLearning}. Ensemble average over 10000 games.}
\label{FIG_SimpleLearningR2}
\end{figure}

Before we proceed in the following section to discuss yet another possibility how to use the compositional memory for learning, it should be noted that many of the observed features can be changed by varying the parameters
$\gamma, R, K$ in the update rules, or by modifying the ways of initializing the memory. For example, as we have seen earlier (in Figure~\ref{FIG_SimpleLearning}), dissipation introduces a mechanism of forgetting, which limits the achievable success probability but at the same time gives the agent more flexibility of adapting to a new strategy of the attacker. To have an agent with \emph{both} a high flexibility \emph{and} a high blocking efficiency, one can choose a finite value of dissipation rate $\gamma$ together with an increased reflection time $R$, as is demonstrated in Figure~\ref{FIG_SimpleLearningR2}. A similar enhancement can be observed for the associativity effect in Figure~\ref{FIG_CompositionAssociation} by increasing $R$ \footnote{For reflection times $R>1$, the emotion tags associated to transitions \tc{$s$} $\to $\textcircled{$a$} must be taken into account. Remember that the emotion tag $\smiley$ (or $\frownie$) associated to the transition \tc{$s$} $\to$ \tc{$a$} is the internal representation of the (most recent) reward $\Lambda(s,a)=1$ (or $0$) assigned to the sequence $(s,a)$. This is also true in the case when the actuator clip \textcircled{$a$} corresponding to actuator $a$ is the result of a deliberation sequence
$\Gamma$ = (\tc{$s$},\tc{$s^{1}$},\dots, \tc{$s^{D}$}, \tc{$a$}) with $D>0$.}.

Another possibility to increase the achievable efficiency is to let the connections of the network dissipate completely when they are not used. While the innate network is characterized by a high connectivity, a trained network will develop both enhanced and suppressed connections.

\subsection{Projective simulation \& learning with composition II}
\lb{ssec:creativity}

In the previous section we saw that projective simulation allowed for associative learning: A novel percept (clip), which had no \emph{a priori} preference for any actuator movement, could excite another clip in episodic memory, from which strong links to specific actuators had been built-up by previous experience. The agent, while presented with a blue arrow, would, with a certain probability, associate it with a red arrow whose meaning it was already familiar with.

A different and more complex behavior can be generated if the agent's actions are not only guided by recalling episodes from the past, but if it can create, as part of the simulation process itself, fictitious episodes that were never perceived before. In the course of the simulation it may for example introduce variations of stored episodes, or it may merge different episodes to a new one, thereby varying or redefining the (virtual) past. The test for all such projections is whether or not the resulting (factual) actions will eventually be rewarded. In other words, it is the performance of the agent in its real life, that \emph{selects} those virtual episodes that have led to successful actions, enhancing the corresponding connections in memory. These principles give the agent a notion of freedom \cite{FreeWill} to ``play around'' with its episodic memories, while at the same time optimizing its performance in the environment.

While it is intuitively clear that such additional capability will be beneficial for the agent, its world (i.e. task environment) must be sufficiently complex to make use of this capability. A typical feature of a complex environment is that the agent can, at some point, ``discover'' new behavioral options that were previously not considered, i.e., not in the standard repertoire of its actions \footnote{For example, when a child learns a certain ability, say, to stand up, for the first time, typically muscles are involved that where never used \emph{in synchrony} (i.e.\ in a specific coordinated way) before, which makes it so difficult. At the same time the realization of \emph{the very possibility} often comes with a strong feeling of pleasure and surprise.}.

\begin{figure}[tb]
\includegraphics[width=8cm]{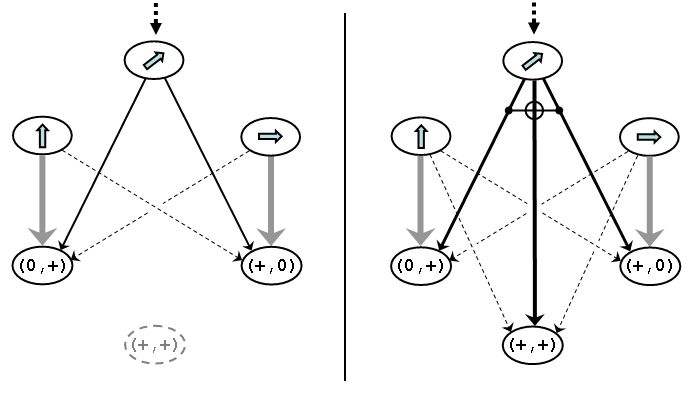}
\caption{Creation of a new and fictitious clip in the memory of the two-dimensional agent. This figure illustrates the schematic evolution of the (relevant part of the) clip network behind Figure \ref{FIG_Creativity}. Frequent excitation of two different actuator clips from a single percept clip leads to the creation of a novel, merged, clip which becomes part of the existing clip network. (See main text.)}
\label{FIG_CreativityMem}
\end{figure}

To map the essential aspects of such a complex situation into our example, we imagine a modification of our invasion game where the defender-agent can move in \emph{two dimensions}, i.e. \texttt{up} and \texttt{down} in addition to \texttt{left} and \texttt{right}. In the notation of Section~\ref{SectionFormalDefinitions}, this corresponds to an enlarged actuator space $A=A_{1}\times A_{2}$ with $a=(a_{1},a_{2})\in \{+,0,-\}\times\{+,0,-\}$ such that, with this notation, \texttt{right}$\equiv (+,0)$, \texttt{left}$\equiv (-,0)$, \texttt{up}$\equiv (0,+)$, \texttt{down}$\equiv (0,-)$. In a robot design, the actuators $a_{1}$ and $a_{2}$ would refer to different motors for motion in $x$ and $y$ direction. One can imagine a two-dimensional array of doors in the $x$-$y$ plane, through which the attacker tries to pass, now entering from the third dimension ($z$-axis). The attacker will move along any of these four directions as well, and use appropriate symbols to announce its moves. However, in addition to those moves, it will at some point start moving also along the \emph{diagonals}, e.g. to the upper-left, in a single step. The defender will first continue to move in the trained directions, simply because the more complex motion along the diagonal is not in its immediate repertoire (although it may technically be able to do it, e.g. by activating the two motors for horizontal and vertical motion at the same time). We assume that there are partial rewards if the defender moves into the right quadrant, e.g.\ by ``blocking'' at least one of the coordinates of the attacker. To be specific, we consider the situation where, from a certain point on, the attacker always moves to the upper-right corner (i.e. along the $+45^{\circ}$ diagonal).
If the agent moves \texttt{right} or \texttt{up}, it will be rewarded, if it moves \texttt{left} or \texttt{down}, it will not. Under the rules specified so far, the agent will, after a transient phase of random motions, be trained so that it will move either up or right, with \emph{equal probability} of $\sim 50\%$ each.
How can the agent conceive of the ``idea'' that it \emph{could} also move along the diagonal direction, by letting both motors run simultaneously, if this composite action was not in its immediate (or: active) repertoire? \footnote{In analogy with the previous example from child development, as described in the preceding footnote, one can imagine that the activation of more complex motions remains suppressed until the basic skills have first been learnt.} The scenario of projective simulation allows for the possibility that, through random clip composition, a merged or mutated clip can be created, that triggers both motors of a composite actuator move. In a sense, the agent would simulate this movement, by chance, before it tries it out in real life. The latter may occur specifically in situations with multiple rewards (or ambivalent moves).

\begin{figure}[tb]
\includegraphics[width=8cm]{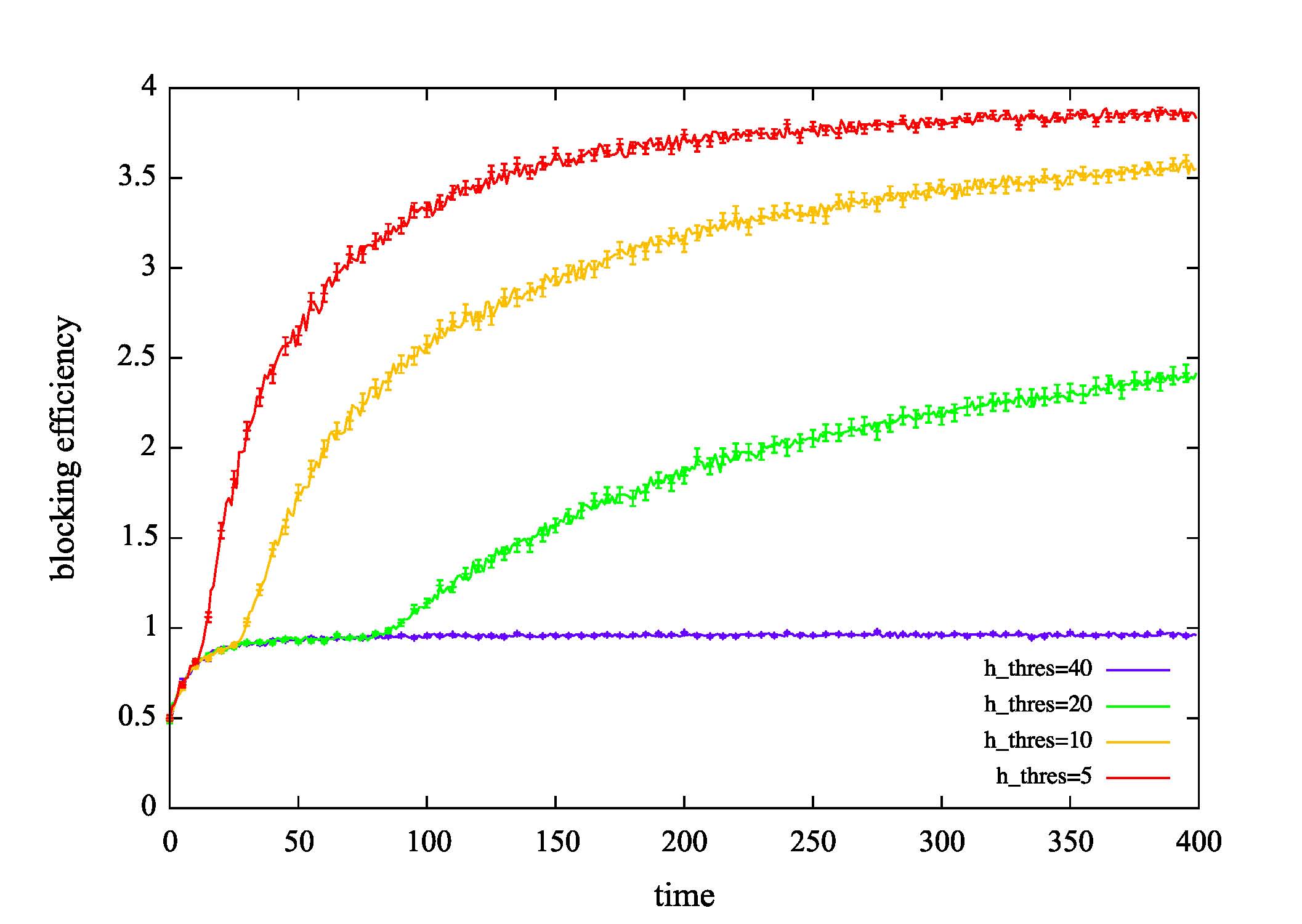}
\caption{Learning curve of a 2D agent (see text) which, after having been trained  on the horizontal and vertical directions (using symbols $\Leftarrow$, $\Rightarrow$ and $\Uparrow$, $\Downarrow$, respectively) is suddenly confronted, at time $n=0$ with moves of the attacker along the diagonal, announced by the symbol {\mbox{\scriptsize $\nearrow$}}. We assume a reinforcement scheme where a movement in the right quadrant (either \texttt{right} or \texttt{up}) is rewarded by a unit increase of the corresponding clip transitions, while a composite movement along the \texttt{diagonal} direction ($+45°$) is rewarded stronger, e.g. by $\lambda =4$. The agent will first quickly learn to move into the right quadrant -- under the rules described in the previous sections -- while on a longer time scale it will find discover the corresponding composite move which the higher reward.}
\label{FIG_Creativity}
\end{figure}

One can think of several possibilities of defining clip merging and variation. A natural possibility exists if, in generalizing our scheme, we allow for parallel excitations of several clips at the same time. Depending on some compatibility constraints, more than one of these clips could then couple out and lead to simultaneous actuator moves.

In the present scenario, however, the simulator can only activate
one clip at a time, but it will happen that two of the clips (e.g.\ those associated to \texttt{right} and \texttt{up}) are activated frequently and with similar probabilities. Here one can e.g.\ define a threshold scheme where a merging of both clips is likely to happen only under the condition that the connections to both of them are sufficiently strong \footnote{Alternatively, one can consider merging of two clips as a second-order process, where it can happen all the time, but with probabilities that are proportional to the product of the individual excitation probabilities.}. The merging itself can be defined on the set of basic elements which make up the clips, obeying certain syntactic constraints.
For example, in the case of the two-dimensional invasion game, we may merge the actuator clips corresponding to \texttt{right} $=(+,0)$ and \texttt{up} $=(0,+)$ into a new clip corresponding to \texttt{right-up} $\equiv (+,+)$, but it is syntactically forbidden to merge \texttt{right} $=(+,0)$ and \texttt{left} $=(-,0)$.

To demonstrate the basic idea, we have implemented a rule according to which the frequent excitation of different actuator clips (of syntactically compatible moves) from a single percept clip creates at some point a novel, merged, actuator clip which becomes part of the clip network. Figure~\ref{FIG_CreativityMem} illustrates the schematic evolution of the (relevant part of the) clip network. The grey arrows indicate previously grown transitions, after the agent has been trained in the horizontal ($\Rightarrow$) and vertical ($\Uparrow$) directions. After such an initial training period, the agent is confronted (dotted arrow) with diagonal moves (see left part of Figure~\ref{FIG_CreativityMem}), announced by the symbol ({\mbox{\scriptsize $\nearrow$}}). When the weights on the two different transitions leaving clip \tc{\mbox{\scriptsize $\nearrow$}} grow beyond a given threshold, a new merged clip is created and connected to \tc{\mbox{\scriptsize $\nearrow$}}, with a weight that is equal to the  sum of the weights on the constitutive transitions. This merging process is indicated schematically in the right part of Figure~\ref{FIG_CreativityMem}.

In Figure~\ref{FIG_Creativity}, we show the resulting learning curve of the agent, which was previously trained ($n<0$, not shown) on the horizontal and vertical directions (using symbols $\Leftarrow$, $\Rightarrow$ and $\Uparrow$, $\Downarrow$, respectively) and is then (at time $n=0$) confronted with moves of the attacker along the diagonal (announced by the symbol ({\mbox{\scriptsize $\nearrow$}}))
\footnote{Actually, the preceding training of the agent on the horizontal and vertical directions is not strictly necessary, in this example, if one assumes that there is an \emph{a priori} connection between the percept clip \tc{\mbox{\tiny $\nearrow$}} and the actuator clips $(+,0)$ and $(0,+)$. Otherwise, the function of the preceding training is to activate those actuator clips for the first time and with it new incoming connections.}. We assume a reinforcement scheme where a movement into the correct quadrant (either \texttt{right} or \texttt{up}) is rewarded by a unit increase of the corresponding weights in the clip network, while a composite movement \texttt{right-up} (both \texttt{right} and \texttt{up}) is rewarded stronger, with $\lambda =4$. One can see that the agent will first quickly learn to move into the right quadrant -- under the rules described in the previous sections -- while on a longer time scale it will discover the corresponding composite move with the higher reward.

%
%

\section{Connection with existing literature}
\label{SectionLiterature}
The problem of learning has been investigated in various fields ranging from psychology, cognitive neuroscience, and philosophy, to artificial intelligence, machine learning, and robotics. In the following, we shall compare our model with some of the works in these fields.

Historically, the idea of using \emph{internal representations} and simulations for learning and prediction was already recognized as a key ingredient for cognitive development in the works by Tolman \cite{Tolman48} (idea of cognitive maps) and Piaget \cite{Piaget71} (role of the internal manipulation of representations). 
The notion of \emph{episodic memory} was introduced in psychology in the 1970s by Tulving \cite{Tulving72} and Ingvar \cite{Ingvar85}, and it has been attracting increasing attention in various fields.
The specific role of episodic memory for simulating future events has recently been discussed by Schacter \emph{et al.} \cite{Schacter08} in the neurosciences, and by Hasselmo \cite{Hasselmo11} who discusses brain mechanisms for episodic memory.

Concepts and ideas for learning play also a major role in artificial intelligence, machine learning and robotics.
The problem of prediction is indeed one of the main topics in machine learning, starting with the seminal work of Holland \cite{Holland75} who introduced the notion of classifier systems, and many subsequent works have used ideas of internal simulation for planning and prediction (for example \cite{Tani96,HoffmannMoeller04,VaughanZuluage06,Toussaint06,ButzEtAl10} and references in reinforcement learning as discussed below). While classifiers \cite{Holland75} bear a certain similarity with the notion of clips that we have introduced in this paper, there are important differences. First, learning classifier systems assume a population or ensemble of classifiers (i.e.\ condition-action rules) and involve a deterministic computation (of the average prediction of a sub-ensemble of classifiers advocating a certain action), after which a specific action is chosen. The random walk through the clip network, in contrast, is much more primitive; it involves no ensemble and no computation. Instead, it amounts to the random hopping through a set of possible clips (including the possibility of creating new clips along the way), without the ability of choosing, sampling, averaging, or in any way optimizing over that set. Every projective simulation corresponds to a single trajectory of a stochastic process (this is important for subsequent quantum generalization, see Section \ref{SectionQuantumAgents}).

In the field of reinforcement learning \cite{SuttonBarto98}, a number of ideas have been discussed which are in some sense related to our work \cite{Sutton90,Lin92,McCallum95,ParrRussel97,SuttonEtAl99,Dietterich00,OrmoneitSen02,SuttonEtAl08}. This concerns in particular the notion of \emph{experience replay} by Lin \cite{Lin92} and recent ideas by Sutton \emph{et al.} \cite{SuttonEtAl08}. 
The work by Lin \cite{Lin92} studies several extensions to standard reinforcement learning algorithms, the most relevant of which, for our present work, is the method of experience replay. In Lin's model, ``by experience replay, the learning agent simply remembers its past experiences and repeatedly presents the experiences to its learning algorithm as if the agent experienced again and again what it had experienced before'' (\cite{Lin92},p.\ 299).
This idea of experience replay has a certain similarity with the our notion of multiple reflections in clip space (indicated by the parameter $R$ in Equation (\ref{CondProb_RRrewarded}) and in Figure~\ref{FIG_ReflectionTimeLearning}); yet, a closer inspection reveals both conceptual and technical differences. The main effect of experience replay in the sense of Lin is to boost the learning process which, in our model, would amount to an (off-line) change of the weights in the clip network. Experience replay is like a module for (self-)teaching: After experiencing a real situation once, the agent gets the chance to review this experience again and again, before taking the next action.
Our notion of episodic memory differs from this one inasmuch as it uses an explicit internal representation and allows more subtle ways of re-using previous experience.
For example, the occurrence of multiple reflections, which also boost the learning speed, is conditioned on the state of certain emotion flags that represent short-time memory. These flags prevent the agent from taking an action that was recently found non-rewarded and give the agent a ``second chance'' to find the right action, but these internal reflections do not change the weights of the clip network. As a second example, the possibility of clip composition (as discussed in Section~\ref{ssec:creativity}) introduces structural changes that also go beyond mere changes of the weights in the clip network. Generally speaking, projective simulation is more integrated with the real actions of the agent; it is a continuous process that runs in parallel (``on-line'') with the real actions.

The work by Sutton \emph{et al.} \cite{Sutton90,SuttonEtAl08} on Dyna-style planning seems in that respect closer to our work. Quoting from Ref.~\cite{SuttonEtAl08}: ``Dyna-style planning proceeds by generating imaginary experience from the world model and then applying model-free reinforcement learning algorithms'', this sounds reminiscent to the use of projective simulation to generate fictitious sequences of memory to guide subsequent action. The underlying conceptual framework is, nevertheless, quite different.
Like most reinforcement learning algorithms, the framework of Dyna-style planning is much more computational than our approach. It uses world models for planning and to decide the course of action. Such planning involves a non-trivial computational process (Dyna-algorithm for policy evaluation) the result of which is then used by the agent to find the optimum course of action. Projective simulation, as mentioned before, is much more primitive; it only involves random hopping through a set of clips, without any further computation. The only parameters that need to be changed and updated in the clip network are the weights of the clip transitions, similar as neural networks (however with the difference that new clips may be created). In that sense, projective simulation is much more embodied and should rather be compared with a biological stochastic process than with the result of planning and computation.

\begin{figure}[H]
\includegraphics[width=8cm]{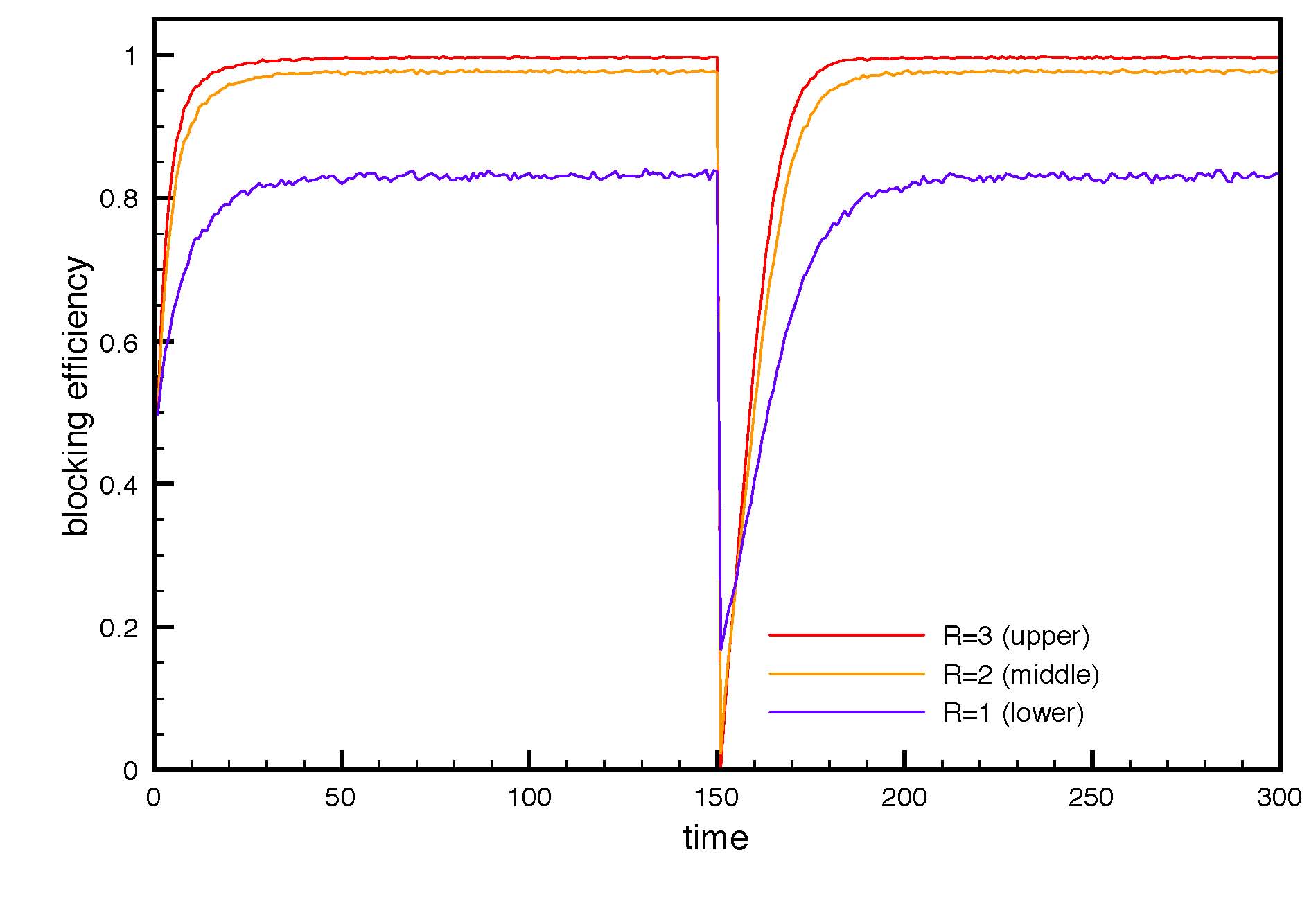}\vspace*{-0.65cm}
\includegraphics[width=8cm]{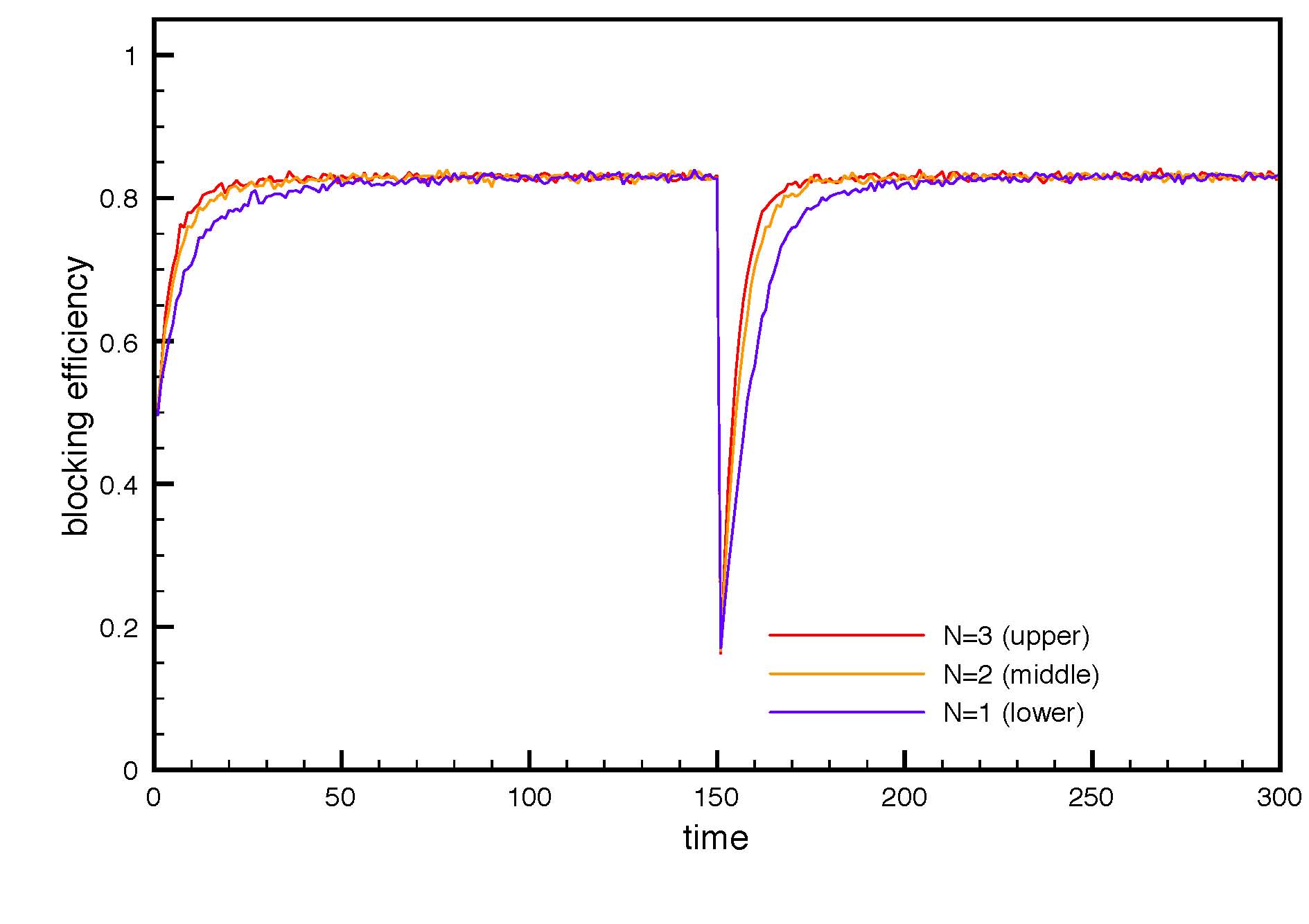}\vspace*{-0.65cm}
\includegraphics[width=8cm]{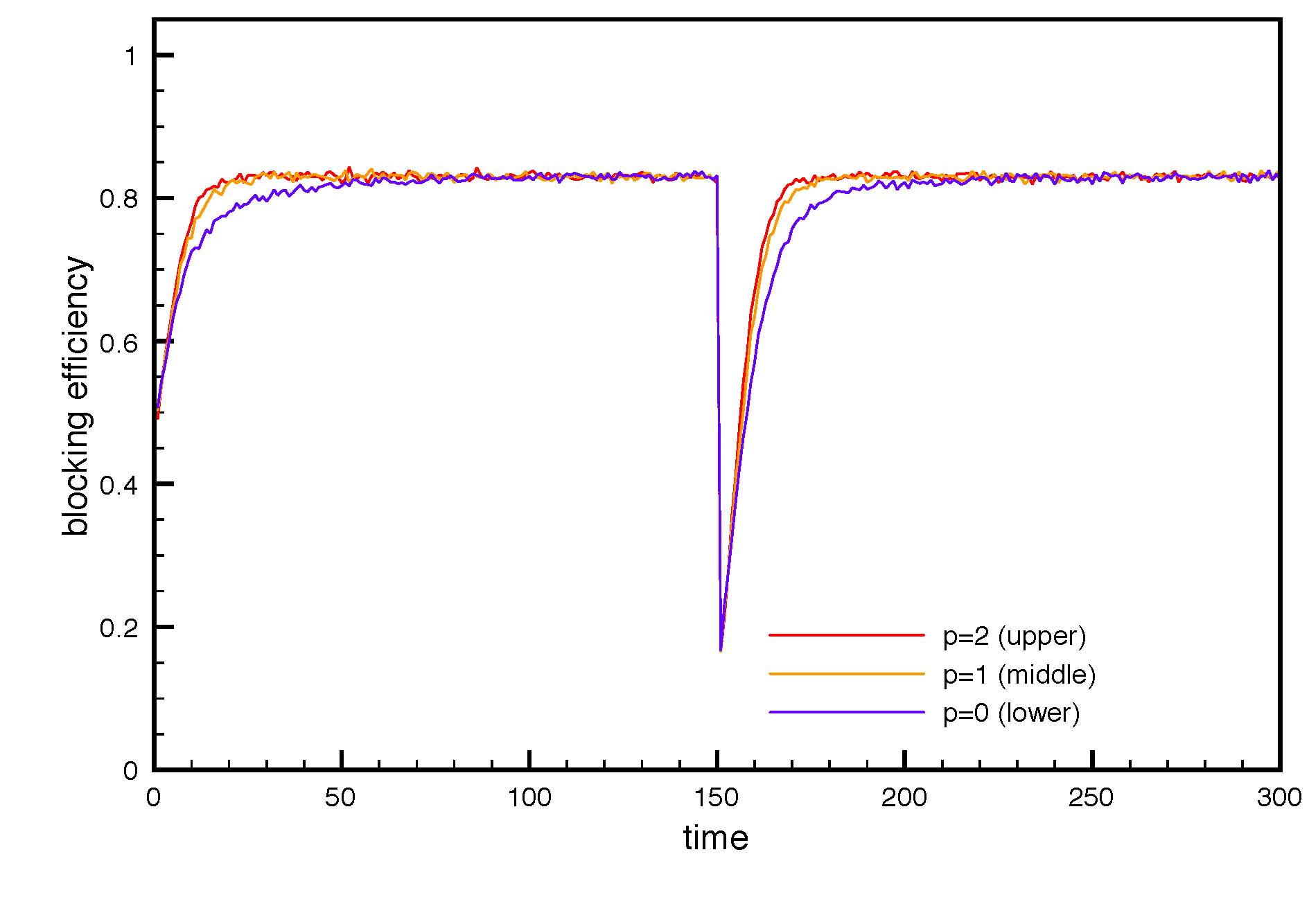}
\caption{
Comparison of projective simulation with experience replay  \cite{Lin92} and Dyna-style planning \cite{SuttonEtAl08}. Learning curves are shown for (a) projective simulation (reflection number $R$) with $\gamma=1/10$ and $\lambda=1$, (b) experience replay (replay number $N$), (c) Dyna-style planning (planning number $p$), whereby both (b) and (c) use the tabular Q-learning algorithm \cite{SuttonBarto98} with a "softmax" action selection rule, based on the Boltzmann distribution. For both (b) and (c) the $Q$ function was initialized to 1, and a reward of 1.5856 was used together with a learning-rate parameter of $\alpha=0.4$. The parameters were chosen such that for $R=1, N=1, p=0$, the initial learning speed and the asymptotic value of the respective learning curves are similar.
In (c) the imagined state and action were picked randomly out of all possible states and actions. It is seen that increasing the parameters $R$, $N$, and $p$ leads to an increased learning speed in each of the respective models, with similar performance. However, different from experience replay and Dyna-style planning, projective simulation with multiple reflection increases not only the learning speed but at the same time the maximum achievable value of the learning parameter (blocking efficiency).
}
\label{FIG_XXXX}
\end{figure}

Despite their conceptual differences, on simple tasks like the invasion game, these different learning models show similar features. In Figure \ref{FIG_XXXX}, we compare the performance of the learning models in the invasion game with two symbols and two actions, $|S|=|A|=2$, where the attacker changes the  meaning of the symbols at $n=150$.  We compare learning curves of  (a) projective simulation, using multiple reflections (reflection number $R$), with (b) experience replay (replay number $N$), and (c) Dyna-style planning (planning number $p$), where the latter two models were based on the Q-learning algorithm \cite{SuttonBarto98}. Increasing the parameters $R$, $N$, and $p$ leads to an
\noindent increased learning speed in each of the respective models, with similar performance. However, different from experience replay and Dyna-style planning, projective simulation with multiple reflections (as defined in Section \ref{SectionProjectedSimulation}) increases not only the learning speed but also the maximum achievable value of the blocking efficiency. The latter can also be achieved in (b) and (c) by changing the external reward.

Generally speaking, we find that on simple tasks like the invasion game the performance of projective simulation is certainly competitive with other modern reinforcement learning algorithms such as experience replay \cite{Lin92} or  Dyna-style planning \cite{SuttonEtAl08}. For more complex task environments these different models may perform differently well on different aspects. With increasing dimension of percept and action space, we find a linear scaling of the learning time with $|S|$ and $|A|$, respectively, similar as for Q-learning \cite{ToBePublished}.
For problems that require long-term planning, we expect methods based on Q-learning or adaptive heuristic critique \cite{SuttonBarto98} to be more favorable, whereas projective simulation with the possibility of clip composition, as discussed in Section \ref{ssec:creativity}, should be favorable in problems  where ``creative'' action in a given situation is in demand. A combination of ideas from projective simulation, such as the use of internal flags encoding short-time memory, with established algorithms for long-time planning is part of an ongoing investigation \cite{ToBePublished}.

\section{Quantum projective simulation}
\label{SectionQuantumAgents}

We now address the generalization of projective simulation to quantum mechanical operation. The motivation of this question is twofold. One reason is the ongoing miniaturization of devices down to the scale of nano-technologies.
It is conceivable that soon robots will be used to control matter even on the molecular and atomic scale, be it in basic research laboratories or in medical applications inside the human body. Agent research will then have to deal with issues of quantum feedback and control \cite{WisemanMilburn09} and its future applications.

Another, more direct, reason has to do with the computational capabilities of quantum computers.
It was found that computers which operate on quantum mechanical principles can solve certain mathematical tasks much more efficiently than any classical computer \cite{NielsenChuang}. It is thus natural to ask whether a similar benefit can be expected for models of artificial intelligence when the architecture of agents involves quantum mechanics.
If one defines an intelligent agent or robot simply as some machine with a ``computer on board'' and with sensors \& actuators as ``input-output devices'', then the answer seems to be straightforward: Replace the classical computer with a quantum computer, run the right quantum algorithm on it, and thus obtain a more efficient agent. The question is then, of course, what is the right quantum algorithm.
A more fundamental problem with this approach is that such a computational viewpoint might miss essential aspects of intelligent \emph{behavior} from the beginning.  It seems that neither a classical computer nor a quantum computer \emph{per se} will make the agent intelligent, nor will any fixed algorithm that runs on these devices. As it has been emphasized in recent literature on artificial intelligence \cite{PfeifferScheier99,FloreanoMattiussi08}, the emergence of intelligent behavior seems to require continuous feedback between the agent and its environment \emph{at its very heart}: In modern terminology,  the agent needs to be \emph{embodied} and \emph{situated} in an environment it interacts with \cite{PfeifferScheier99}. Modern notions of (reinforcement) learning and agents are developed within this framework, and so is our approach to creative behavior, in which the network of clips i.e. the episodic memory \emph{grows as the agent interacts} with the world. Furthermore, the evolution of the episodic memory (clip network) is thereby firmly embedded in the agent architecture.

In the following we describe how the model of projective simulation can be generalized in the quantum regime, introducing a notion of quantum agents. In quantum mechanics, states of a system are described by vectors (or rays) in a complex Hilbert space, and observables by linear Hermitean operators acting on that space. A quantum-enhanced autonomous agent can be defined as an agent that interacts with a classical environment, but whose memory (or, more generally, internal state) uses quantum degrees of freedom \footnote{There are other situations conceivable where the environment is quantum mechanical, and the task of the agent is to bring the environment into a certain quantum state. Depending on whether the agent employs quantum degrees of freedom itself -- in its memory, its sensors, or its actuators -- one can define a variety of different agents.}. In the notation and terminology of Section \ref{SectionFormalDefinitions}, the external variables $s$ (percepts) and $a$ (actions) are then still classical variables, while the clips $c\in C$ become quantum states $|c\rangle \in H_C$ (Hilbert space of the memory). An external stimulus $s$ will excite memory in a quantum state $|c\rangle = |$\tc{$s$}$\rangle$ (the percept clip) which has now the status of a basis state in the memory system. The random walk in clip space, which is an essential ingredient in our model, now becomes a \emph{quantum walk} in the associated Hilbert space of the (quantum) memory, with the replacements
\begin{equation}
p(c'|c)\longrightarrow |\langle c'|c\rangle |^2
\end{equation}
for elementary transitions between clips, and
\begin{equation}
p(c''|c) \longrightarrow |\sum_{c'}\langle c''|c' \rangle\langle c'|c \rangle|^2.
\end{equation}
for composite transitions. Here the scalar product $\langle c'|c\rangle$ defines the \emph{probability amplitude} for the transition \tc{$c$}$\to$\tc{$c'$}, and the modulus squared in the expression for the composite transition gives rise to \emph{quantum interference}, which is one of the basic features of quantum mechanics. Quantum interference is in particular exploited in fast algorithms for quantum search \cite{Grover97} and quantum walks on graphs \cite{Aharonov01}.

Let us now describe the quantization procedure in more detail. With the clip network as illustrated in Figure \ref{ClipNetwork} one can associate a graph $G=(V,E)$, where the vertices $j\in V$ label the different clips $c_j\in C$ within the network and the edges $\{j,k\}\in E$ denote possible transitions between clips. A quantum walk in memory space is then generated by a Hamiltonian of the form \cite{Stamp07}
\begin{eqnarray}\label{Hamiltonian}
H &=& \sum_{\{j,k\}\in E} \lambda_{jk}\left(\hat c_k^{\dag}\hat c_j + \hat c_k\hat c_j^{\dag}\right) + \sum_{j\in V} \epsilon_{j} \hat c_j^{\dag} \hat c_j \\
& = & \sum_{\{j,k\}\in E} \lambda_{jk}\left(|c_k\rangle\langle c_j| + |c_j\rangle\langle c_k|\right) + \sum_{j\in V} \epsilon_{j} |c_j\rangle\langle c_j| \nonumber
\end{eqnarray}
where the operator $\hat c_{j}$ excites the memory from its ground state into clip $c_j$,
\begin{equation}
| c_j \rangle = \hat c^{\dag}_{j}|\texttt{vac}\rangle
\end{equation}
and $\hat c_k^{\dag} \hat c_j$ induces a transition $c_j \rightarrow c_k$:
\begin{equation}
\hat c_k^{\dag}\hat c_{j}|c_j\rangle = |c_k\rangle .
\end{equation}

The dynamical equation that describes the coherent quantum walk is given by the Liouville-von Neumann equation
\begin{equation}
\frac{\partial}{\partial t} \rho = -i [H, \rho]
\label{VonNeumann}
\end{equation}
where $\rho=\rho(t)$ is the quantum state (density operator) of the memory at time $t$, $[H, \rho]\equiv H \rho - \rho H$ is the commutator, and we have set Planck's constant to unity.

The (real) coupling parameters $\lambda_{jk}$ in (\ref{Hamiltonian}) induce coherent transitions between the different clips in the network.
One can also include further, incoherent, transitions described by a Liouvillean operator of the type
\begin{equation}
L \rho = \sum_{\{j,k\}\in E} \kappa_{jk}\left( \hat c_k^{\dag} \hat c_j \rho \hat c_k \hat c_j^{\dag} - \frac{1}{2}\{\hat c_k^{\dag} \hat c_j\hat c_k \hat c_j^{\dag} \rho + \rho \hat c_k^{\dag} \hat c_j\hat c_k \hat c_j^{\dag}\}\right)
\label{Lindblad}
\end{equation}
with $\kappa_{jk}\ge 0$, in which case (\ref{VonNeumann}) generalizes to the \emph{quantum master equation}
\begin{equation}
\frac{\partial}{\partial t} \rho = -i [H, \rho] + L \rho .
\label{MasterEquation}
\end{equation}

The dynamical equation (\ref{MasterEquation}) represents a generalization to the master equation/stochastic process that describes the classical random walk, which is formally recovered in the limit where $H=0$.
The transitions generated by the Hamiltonian part are coherent and give rise to quantum superpositions and interference, which lies at the heart of the \emph{quantum parallelism} that is exploited in quantum computers and in quantum walks. The incoherent transition generated by the Lindblad part can be interpreted as the result of spontaneous ``quantum jumps'' between different clip states.

Most examples of quantum walks that have been studied correspond to walks on undirected graphs. A possibility to introduce directed walks is to add incoherent transitions generated by (\ref{Lindblad}). The price one has to pay with such directed transition is that they introduce decoherence, so in general there will be a balance between quantum coherence on one side, and directedness on the other side. In combining these elements, one can design walks with coherent, bi-directional transitions in certain regions of the network (or graph), combined with incoherent transitions that ``project'' to other regions, or that exit the clip network. The Hamiltonian used in (\ref{Hamiltonian}) can be generalized to so-called composite walks \cite{Stamp07} that include further degrees of freedom associated with a given transition, which could be used to include the emotion tags (see Section \ref{SectionFormalDefinitions}) into the quantum mode, as well as to implement discrete quantum walks using quantum coins \cite{Kempe03}.

The clips themselves have a composite structure and may include remembered percepts $s\in S$ or actions $a\in A$, each of which can be composed of different categories. This compositional structure is accounted for by a tensor-product in the Hilbert space of the clips. For example, in case of a percept clip
$c=\mu(s)$, the corresponding clip operators have the form
\begin{equation}
\hat c^{\dag} = \hat\mu^{\dag}(s) = \hat\mu^{\dag}_1(s_1)\otimes\hat\mu^{\dag}_2(s_2)\dots \hat\mu^{\dag}_N(s_N)
\end{equation}
where $\hat\mu^{\dag}_i$ is the memory operator that excites percept of category $i$ (like, for example, color or shape).

A call of episodic memory in this picture involves three steps, which also illustrates the embedding of the quantum walk into the otherwise
classical agent architecture:

\begin{itemize}
\item Memory activation. Classical percept $s\in S$ triggers the excitation of an associated memory state: $s\mapsto \rho(s)=|\psi(s)\rangle\langle \psi(s)|$. {\small (In the simplest case, $|\psi(s)\rangle = |s\rangle  = \hat\mu^{\dag}(s)|\texttt{vac}\rangle$, but $|\psi(s)\rangle$ could also involve superpositions of several percept states related to $s$.)} \newline
\item Quantum walk through the network of clips, as described by the quantum master equation (\ref{MasterEquation}) with Hamiltonian (\ref{Hamiltonian}) and with $|\psi(s)\rangle$ as initial state. \newline
\item Memory output. A classical signal that induces (real) action is generated by the measurement of certain memory observables. {(\small In the examples described in Section \ref{SectionSimpleExamples} these are the actuator observables $\hat\mu^{\dag}(a)\hat\mu(a)$, and the probability $p_t(a)$ for an actuator motion $a$ to be triggered at time $t$ is given by $p_t(a)={\rm tr}(\hat\mu^{\dag}(a)\hat\mu(a)\rho(t))= {\rm tr}(\hat\mu(a)\rho(t)\hat\mu^{\dag}(a))$ where $\rho(t)$ is the state of the memory at time $t$.)}
\end{itemize}

This described model represents a generalization of the classical random walk, which can be recovered from (\ref{MasterEquation}) by switching off the coherent interactions. It is clear that the possibility of creating quantum superpositions of many different percept states opens the door for potentially huge speed-ups in exploring memory \cite{Kempe03}, which is subject of an ongoing investigation \cite{ToBePublished}. Note that quantum random walk processes similar to (\ref{MasterEquation}), with engineered quantum many-body interactions, have recently been realized in the context of dissipation-driven quantum simulation with trapped ions \cite{Barreiro11}. Similarly, quantum simulators based on laser-driven atomic gases in optical lattices have been proposed \cite{WeimerEtAl10,Diehl08} and are currently being explored in many laboratories.

The scheme that we have presented can be extended into various ways. Instead of a simple quantum walk, one can also introduce additional quantum computational elements when calling and processing episodes in memory space.
A more detailed exposition of these ideas is beyond the scope of this paper and will be given in future work.

\section{Conclusion}
\label{SectionConclusion}

We have introduced the notion of projective simulation and discussed its potential role for learning in artificial agents. We have shown that it allows an agent to project itself into fictitious situations, which are self-generated by the agent (and its specific memory system) and which influence its future actions. Projective simulation enhances the learning capabilities of an agent and introduces an elementary notion of creative action. To illustrate the basic concepts, we have worked out simple but concrete examples of learning agents and the interplay of simulation and episodic memory (ECM). We have programmed a learning agent that uses projective simulation, studied its behavior and tested its performance in the invasion game. The idea of projective simulation is however more general and we believe that the scheme, as part of a comprehensive embodied approach to artificial intelligence, could be implemented in autonomous agents or robots with realistic task environments.

We believe that the ``embodied approach'' to artificial intelligence parallels in some way the recent strong attention to the role of physics for the foundations of computer science (down to the level of quantum mechanics).
In a similar spirit as people have studied the ultimate power of computers on the basis of physical law \cite{Feynman82,Deutsch85}, we are here concerned with the question of the ultimate scope of intelligent behavior in embodied agents, taking  into account the physical basis of this embodiment. To approach this question, one first needs to develop a model of simulation in agents that is both physically grounded and at the same time general in its constitutive concepts (i.e. not linked to a specific implementation). We have shown that the abstract notion of clips and of projected simulation as a random walk through the space of clips, which grows dynamically by the specified rules of clip variation and composition, provides a first step towards such a general framework. From a physicist's perspective, such a random walk can be understood as the propagation of excitations of physical degrees of freedom that represent the information carrying quantities. Within such conceptual framework, we can formulate, for the first time, a meaningful notion of an embodied quantum agent, by extending the model of projective simulation to the quantum regime.

\emph{Acknowledgements:}
We thank Julian Mautner, Adi Makmal, and Daniel Manzano for discussions and for leaving us
part of their numerical results prior to publication. The work was supported by the Austrian
Science Fund (FWF) through projects F04011 and F04012.



\begin{thebibliography}{99}


\bibitem{SuttonBarto98}
Sutton, Richard S. \& Barto, Andrew G.
\newblock \emph{Reinforcement learning}.
\newblock First edition (MIT Press, Cambridge Massachusetts, 1998).

\bibitem{RusselNorvig03}
Russel, Stuart J. \& Norvig, Peter.
\newblock \emph{Artifical intelligence - A modern approach}.
\newblock Second edition (Prentice Hall, New Jersey, 2003).

\bibitem{PfeifferScheier99}
Pfeiffer, Rolf \& Scheier, Christian.
\newblock \emph{Understanding intelligence}.
\newblock First edition (MIT Press, Cambridge Massachusetts, 1999).

\bibitem{FreeWill}
Briegel, Hans J.
\newblock On creative machines and the physical origins of freedom.
\newblock \emph{Sci. Rep.} \textbf{2}, 522 (2012)

\bibitem{NielsenChuang}
M.A. Nielsen and I.L. Chuang.
\newblock Quantum computation and quantum information,
\newblock First edition (Cambridge University Press, Cambridge 2000).

%

\bibitem{FloreanoMattiussi08}
Floreano, Dario \& Mattiussi, Claudio.
\newblock \emph{Bio-inspired artificial intelligence : theories, methods, and technologies}.
\newblock First edition (MIT Press, Cambridge Massachusetts, 2008).

\bibitem{Tulving72}
Tulving, Endel.
\newblock Episodic and semantic memory.
\newblock In \emph{Organization of Memory}, ed. E Tulving,
W Donaldson, pp. 2381--403 (1972).
For a recent review see Tulving, Endel,
\newblock Episodic memory: From mind to brain,
\newblock \emph{Annu. Rev. Psychol.} \textbf{53},  1-25 (2002).

\bibitem{Ingvar85}
Ingvar, D. H.
\newblock ``Memory of the future'': An essay on the temporal organization of conscious awareness.
\newblock \emph{Human neurobiology} \textbf{4} 127-136 (1985).

\bibitem{Tolman48}
Tolman, Edward C.
\newblock Cognitive maps in rats and men.
\newblock \emph{The Psychological Review} \textbf{55(4)}, 189-208 (1948)

\bibitem{Piaget71}
Piaget, Jean.
\newblock \emph{Mental imagery in the child: a study of the development of imaginal representation}.
\newblock (London: Routledge and Kegan Paul, 1971).

\bibitem{Heidegger26}
Heidegger, Martin,
\newblock \emph{Sein und Zeit},
\newblock Sixteenth edition (Max Niemeyer Verlag, T\"ubingen, 1986).
\newblock English translation: \emph{Being and Time} (translated by John Macquarrie and Edward Robinson)
\newblock First edition (Blackwell Publishing, 1962).

\bibitem{ClarkGrush99}
Clark, Andy. \& Grush, Rick.
\newblock Towards a Cognitive Robotics.
\newblock \emph{Adaptive Behavior} \textbf{7}, 5-16, (1999)

\bibitem{Dennett91}
Dennett, Daniel C.,
\newblock \emph{Consciousness explained.}
\newblock First paperback edition (Bay Back Books, Boston, 1991).

\bibitem{Hesslow02}
Hesslow, Germund.
\newblock Conscious thought as simulation of behaviour and perception.
\newblock \emph{TRENDS in Cognitive Sciences} \textbf{6}, 242-247, (2002).

\bibitem{Schacter08}
Schacter, Daniel L., Addis, Donna Rose \& Buckner, Randy L.
\newblock Episodic Simulation of Future Events: Concepts, Data, and Applications.
\newblock \emph{Ann. N.Y. Acad. Sci.} \textbf{1124}, 39–60, (2008).

\bibitem{Hasselmo11}
Hasselmo, Michael E.,
\newblock \emph{How we remember. Brain mechanisms of episodic memory.}
\newblock First edition (MIT Press, Cambridge Massachusetts, 2011).


\bibitem{Lin92}
Lin, Long-Ji.
\newblock Self-improving reactive agents based on reinforcement learning, planning and
teaching.
\newblock \emph{Machine Learning} \textbf{8}, 292-321 (1992).

\bibitem{Sutton90}
Sutton, R. S.
\newblock Integrated architectures for learning, planning, and reacting based on approximating dynamic programming.
\newblock \emph{Proceedings of the Seventh International Conference on Machine Learning ICML'90}, Morgan Kaufmann, pp.\ 216-224 (1990).

\bibitem{SuttonEtAl99}
Sutton, R.S., Precup, D., Singh, S.
\newblock Between MDPs and semi-MDPs: A Framework for Temporal Abstraction in Reinforcement Learning.
\newblock \emph{Artificial Intelligence}, \textbf{112}, 181-211 (1999).

\bibitem{OrmoneitSen02}
Ormoneit, D. \& Sen, S.
\newblock Kernel-based reinforcement learning.
\newblock \emph{Machine Learning}, \textbf{49}, 161–178 (2002)

\bibitem{SuttonEtAl08}
Sutton, R. S., Szepesvari, Cs., Geramifard, A. \& Bowling, M.
\newblock Dyna-style planning with linear function approximation and prioritized sweeping.
\newblock \emph{Proceedings of the 24th Conference on Uncertainty in Artificial Intelligence}, pp.\ 528-536 (2008).

\bibitem{McCallum95}
McCallum, R. Andrew,
\newblock  Instance-Based Utile Distinctions for Reinforcement Learning with Hidden State.
\newblock \emph{Proceedings of the Twelfth International Conference on Machine Learning}, Morgan Kaufmann, pp.\ 387-395 (1995).

\bibitem{ParrRussel97}
Parr, R and Russell, S.
\newblock Reinforcement learning with hierarchies of machines.
\newblock \emph{NIPS}, \textbf{10}, 1043-1049 (1998).

\bibitem{Dietterich00}
Dietterich, T. G.
\newblock Hierarchical reinforcement learning with the MAXQ value function decomposition.
\newblock \emph{Journal of Artificial Intelligence Research}, \textbf{13}, 227-303 (2000).

\bibitem{Tani96}
Tani, Jun.
\newblock Model-Based Learning for Mobile Robot Navigation from the Dynamical Systems Perspective.
\newblock \emph{IEEE Trans. System, Man and Cybernetics}, \textbf{26}, 421-436 (1996).

\bibitem{HoffmannMoeller04}
Hoffman, H. \& M{\"o}ller, R.
\newblock Action Selection and Mental Transformation Based on a Chain of Forward Models.
\newblock In Schaal \emph{et al.} (eds.) \emph{Proceedings of the 8th Conference on Simulation of Adaptive Behavior (SAB '04)}, pp.\ 213-222, MIT Press (2004)

\bibitem{VaughanZuluage06}
Vaughan, R. \& and Zuluaga, M.
\newblock Use your illusion: Sensorimotor Self-simulation allows complex agents to plan with incomplete self-knowledge.
\newblock In: Nolfi \emph{et al.} (eds.) SAB 2006, LNCS (LNAI) \textbf{4095}, 298-309, Springer (2006)

\bibitem{Toussaint06}
Toussaint, M.
\newblock A sensorimotor map: Modulating lateral interactions for anticipation and planning.
\newblock \emph{Neural Computation} \textbf{18}, 1132-1155 (2006).

\bibitem{ButzEtAl10}
Butz, Martin V., Shirinov, Elshad \& Reif Kevin L.
\newblock Self-Organizing Sensorimotor Maps Plus Internal Motivations Yield Animal-Like Behavior.
\newblock \emph{Adaptive Behavior}, \textbf{18}, 315-337 (2010).

\bibitem{Holland75}
Holland, John H.
\newblock \emph{Adaptation in natural and artificial systems: an introductory analysis with applications to biology, control, and artificial intelligence}.
\newblock (University of Michigan Press, Ann Arbor 1975).

\bibitem{Watkins89}
Watkins, C.J.C.H.
\newblock \emph{Learning from delayed rewards}.
\newblock PhD Thesis, University of Cambridge, England, 1989.



\bibitem{Braitenberg86}
Braitenberg, Valentino.
\newblock \emph{Vehicles: Experiments in synthetic psychology}.
\newblock First paperback edition (MIT Press, Cambridge Massachusetts, 1986).


\bibitem{KandelNobelPrize}
Kandel, Eric.
\newblock
The molecular biology of memory storage: A dialog between genes and synapses.
\newblock
in ``Nobel Lectures, Physiology or Medicine 1996-2000,'' Editor Hans Jörnvall (World Scientific Publishing Co., Singapore, 2003).

\bibitem{KandelAntonov03}
Antonov, Igor; Antonova, Irina; Kandel, Eric R. \& Hawkins, Robert D.
\newblock Activity-dependent presynaptic facilitation and Hebbian LTP are both required and interact during classical conditioning in Aplysia.
\newblock \emph{Neuron} \textbf{37} (1), 135–147 (2003).

\bibitem{MartinHeisenberg}
Heisenberg, Martin, \emph{et al.}
\newblock Attracting a fly's attention.
\newblock Invited Talk at the ESF-EMBO Conference
\emph{Functional Neurobiology in Minibrains: From Flies to Robots and Back Again},
17-22 October 2010, Sant Feliu de Guixols, Spain.

\bibitem{HeisenbergPNAS}
Sareen, Preeti S.; Wolf, Reinhard;
\& Heisenberg, Martin.
\newblock Attracting the attention of a fly.
\newblock \emph{PNAS} \textbf{108}, 7230--7235 (2011).

\bibitem{ToBePublished}
Mautner, J., Makmal, A., \emph{et al.}, unpublished manuscript (2012).

\bibitem{GodsilRoyle01}
Godsil, Chris \& Royle, Gordon.
\newblock \emph{Algebraic Graph Theory.}
\newblock First edition (Springer, New York, 2001).

\bibitem{WisemanMilburn09}
Wiseman, H. M. \& Milburn, G. J.
\newblock \emph{Quantum Measurement and Control}.
\newblock (Cambridge University Press, 2009)

\bibitem{Grover97}
Grover, Lev.
\newblock Quantum Mechanics helps in searching for a needle in a haystack.
\newblock \emph{Physical Review Letters} \textbf{79}, 325-328, (1997).

\bibitem{Aharonov01}
Aharonov, Dorit; Ambainis, Andris; Kempe, Julia \& Vazirani, Umesh.
\newblock Quantum Walks On Graphs.
\newblock Proceedings of ACM Symposium on Theory of Computation
(STOC'01), July 2001, pp. 50-59.

\bibitem{Stamp07}
Hines, A. P. \& Stamp, P.C.E.,
\newblock Quantum walks, quantum gates, and quantum computers.
\newblock \emph{Physical Review A} \textbf{75}, 062321 (2007).

\bibitem{Kempe03}
Kempe, Julia
\newblock Quantum random walks - an introductory overview.
\newblock \emph{Contemporary Physics} \textbf{44}, 307–327 (2003).

\bibitem{Barreiro11}
Barreiro J.T. \emph{et al.}
\newblock An open-system quantum simulator with trapped ions.
\newblock \emph{Nature} \textbf{470}, 486-491 (2011).

\bibitem{WeimerEtAl10}
Weimer, Hendrik, \emph{et al.}
\newblock A Rydberg quantum simulator.
\newblock \emph{Nature Physics} \textbf{6}, 382-388 (2010).

\bibitem{Diehl08}
Diehl, Sebastian \emph{et al.}
\newblock Quantum states and phases in driven open quatum systems with cold atoms.
\newblock \emph{Nature Physics} \textbf{4}, 878-883 (2008).

\bibitem{Feynman82}
Feynman, Richard.
\newblock Simulating physics with computers.
\newblock \emph{Int. J. Theor. Phys.} \textbf{21} 467-488 (1982).

\bibitem{Deutsch85}
Deutsch, David
\newblock Quantum Theory, the Church-Turing Principle and the Universal Quantum Computer.
\newblock \emph{Proc. R. Soc. Lond. A} \textbf{400} 97-117 (1985).

\end{thebibliography}
\end{document}